\documentclass[acmsmall,authorversion,nonacm]{acmart}
\makeatletter
\AtBeginDocument{\let\hl\@firstofone}
\makeatother
\AtBeginDocument{%
  \providecommand\BibTeX{{%
    \normalfont B\kern-0.5em{\scshape i\kern-0.25em b}\kern-0.8em\TeX}}}

\usepackage{float}
\usepackage{makecell}
\usepackage{soul}
\usepackage{import}
\usepackage{tcolorbox}
\usepackage{glossaries}
\usepackage{comment}
\usepackage{pdfpages} 
\usepackage{comment}
\usepackage{array}

\usepackage{longtable}
\usepackage{xcolor}
\usepackage{comment}

\usepackage{xcolor,colortbl}
\definecolor{Gray}{gray}{0.85}
\newcolumntype{C}{>{\centering\let\newline\\\arraybackslash\hspace{0pt}}m{0.2\textwidth}}
\newcolumntype{A}{>{\centering\let\newline\\\arraybackslash\hspace{0pt}}m{0.05\textwidth}}
\newcolumntype{V}{>{\centering\let\newline\\\arraybackslash\hspace{0pt}}m{0.4\textwidth}}
\newcolumntype{L}{>{\let\newline\\\arraybackslash\hspace{0pt}}m{0.75\textwidth}}
\newcolumntype{B}{>{\let\newline\\\arraybackslash\hspace{0pt}}m{0.82\textwidth}}
\newcolumntype{M}{>{\let\newline\\\arraybackslash\hspace{0pt}}m{0.1\textwidth}}

\newcolumntype{A}[1]{>{\raggedright\let\newline\\\arraybackslash\hspace{0pt}}m{#1}}

\newcolumntype{B}[1]{>{\centering\let\newline\\\arraybackslash\hspace{0pt}}m{#1}}

\usepackage{graphicx}

\begin{document}

\title{Robots as Mental Well-being Coaches: Design and Ethical Recommendations}

\author{Minja Axelsson}
\email{minja.axelsson@cl.cam.ac.uk}
\orcid{0000-0002-1101-2539}
\affiliation{%
  \institution{AFAR Lab (Affective Intelligence and Robotics Laboratory); Department of Computer Science and Technology; University of Cambridge}
  \city{Cambridge}
  \country{United Kingdom}
}

\author{Micol Spitale}
\email{micol.spitale@polimi.it}
\orcid{0000-0002-3418-1933}
\authornote{This work was undertaken and finalised while Micol Spitale was a postdoctoral researcher at the University of Cambridge.}
\affiliation{%
  \institution{Politecnico di Milano}
  \city{Milan}
  \country{Italy}
}

\author{Hatice Gunes}
\email{hatice.gunes@cl.cam.ac.uk}
\orcid{0000-0003-2407-3012}
\affiliation{%
  \institution{AFAR Lab (Affective Intelligence and Robotics Laboratory); Department of Computer Science and Technology; University of Cambridge}
  \city{Cambridge}
  \country{United Kingdom}
}

\renewcommand{\shortauthors}{Axelsson, et al.}

\begin{abstract}
The last decade has shown a growing interest in robots as well-being coaches. However, insightful guidelines for the design of robots as coaches to promote mental well-being have not yet been proposed. This paper details design and ethical recommendations based on a qualitative analysis drawing on a grounded theory approach, which was conducted with a three-step iterative design process which included user-centered design studies involving robotic well-being coaches, namely: (1) a user-centred design study conducted with 11 participants consisting of both prospective users who had participated in a Brief Solution-Focused Practice study with a human coach, as well as coaches of different disciplines, (2) semi-structured individual interview data gathered from 20 participants attending a Positive Psychology intervention study with the robotic well-being coach Pepper, and (3) a user-centred design study conducted with 3 participants of the Positive Psychology study as well as 2 relevant well-being coaches. After conducting a thematic analysis and a qualitative analysis, we collated the data gathered into convergent and divergent themes, and we distilled from those results a set of design guidelines and ethical considerations. Our findings can inform researchers and roboticists on the key aspects to take into account when designing robotic mental well-being coaches.

\end{abstract}

\ccsdesc[500]{Computer systems organization~Robotics}
\ccsdesc[300]{Human-centered computing~HCI design and evaluation methods}
\ccsdesc[100]{Human-centered computing~User studies}
\ccsdesc[500]{Human-centered computing~User centered design}
\ccsdesc[300]{Human-centered computing~HCI theory, concepts and models}
\ccsdesc[300]{Computing methodologies~Artificial intelligence}

\keywords{Human-Robot Interaction, Social Robots, Robot Design, Design Guidelines, Ethical Considerations}

\maketitle

\newpage

\section{Introduction}

The prevalence of mental health conditions has grown significantly in the last decade -- with an increase of 13\% in people diagnosed with mental health disorders \cite{who2018} --  \hl{and it has been even more exacerbated by the COVID-19 pandemic. This increase is due to several risk factors, such as the use of social media, the inequality and social disadvantage, discrimination and social exclusion, traumatic experiences, differences in physical health etc.} \cite{sheldon2021prevalence, giacco2018prevalence}.
These conditions can have a negative impact on people's everyday life, affecting for example their academic or professional achievements, relations with family and friends, and their community involvement. 
Despite these numbers, less than 2\% of worldwide government health spending is allocated to mental health \hl{even if more would be needed to meet the request of mental health services}.
The limitations in accessibility of mental well-being practices, the lack of personnel and resources to support people in this regard, and the advent of mental health digital applications (e.g., mobile apps) --- widely used especially during the COVID-19 pandemic \cite{alexopoulos2020use, blake2020mitigating} --- have contributed to the increase of \hl{research} interest in the use of robots as coaches for well-being \cite{spitale2022affective}.

\hl{To respond to those requests, many tech companies offered affordable and accessible mental well-being service, such as apps (e.g., HeadSpace, Calm) that have been shown to have a great potential to help the promotion of people's mental well-being }\cite{spitale2023, axelsson2023robotic}. \hl{Despite this, recent studies} \cite{torous2020dropout, kerst2020smartphone} \hl{indicated that participants struggled to find time to use a meditation app, and that a high dropout rate and low app engagement represent well-known hurdles} \cite{laurie2016making}.
\hl{On the contrary, past works in Human-Robot Interaction (HRI) have showed that robots can help in promoting mental health application in the long term, such as for mindfulness exercises }\cite{bodala2021teleoperated} or positive psychology practices \cite{spitale2023}, as well as supporting physical well-being, such as in rehabilitative physical therapy \cite{winkle2018social}, even if they may be expensive with respect to app-based solutions. 
\hl{Previous works have investigated socially assistive robots from the perspective of well-being, care, and justice, and has called for the formulation of practical principles for SAR design} \cite{boada2021ethical}. As explorations of robots as mental well-being coaches are still quite novel, \hl{cohesive guidelines for their design do not exist}. Guidelines have been previously presented to share knowledge on how to design robots for specific applications, such as for communication therapy for children with autism \cite{axelsson2019participatory}, and for encouraging children's creativity \cite{alves2021children}.
In this paper, we aim to present such design guidelines for robotic mental well-being coaches, to further inform the field on how to design them.

To this end, we have \hl{followed an iterative design process} 
\hl{by conducting} three different user-centred design studies exploring a robotic mental well-being coach. In the first study, we involved 11 participants -- including University's students --who had previous experience with well-being practices.
We collected their feedback for a hypothetical robot (i.e., participants were not asked to think about a specific, pre-existing robot), gathering a broad range of user perspectives and needs on different types of well-being practices a robotic coach could conduct. In the second study, building upon the first study results, we conducted interviews with 20 participants -- recruited among the University's students -- who had just interacted with a robotic well-being coach conducting positive psychology sessions. 
In the third study, 5 participants (both prospective users and coaches) were invited to have a larger group discussion based on their experience in the previous study, and then to explicitly edit robotic coach dialogue.

We analyzed the data collected with Thematic Analyis (TA) \cite{braun2006using}, and then we conducted a \textcolor{black}{qualitative analysis inspired by the} Qualitative Meta-Analysis method \cite{timulak2014qualitative} of the three studies. The results included what advantages (e.g., improved accessibility and lack of judgement) and disadvantages (e.g., privacy concerns and lack of humanness) prospective users expect to have in a potential robotic mental well-being coach. Additionally, participants detailed features and capabilities they would expect to see in such a robot. 
Then, we operationalized those results as recommendations and guidelines, in order to contribute to the future design of such robots. 
The major contributions of this paper are the following:
\begin{itemize}
    \item we report results from three user-centered design studies highlighting the design features and capabilities that a robot should have to be used as a well-being coach;
    \item based on the lessons learned from the studies, we identified a set of recommendations in terms of design guidelines and ethical considerations to inform researchers and roboticists about the key aspects to take into account during the design process of a robotic coach for mental well-being.
\end{itemize}

The rest of the paper is organized as follows. Section \ref{sec:rw} summarizes the background and related works that informed this research, Section \ref{sec:methodology} describes the methodology for our research, Section \ref{sec:study1},  \ref{sec:study2} and \ref{sec:study3} describe respectively Study 1, Study 2 and Study 3. Section \ref{sec:meta-analysis} reports the \textcolor{black}{qualitative} analysis from the results of the previous three studies, Section \ref{sec:design} reports the design guidelines obtained from the studies, and discusses ethical consideration, and Section \ref{sec:concl} concludes the paper and proposes directions for future research.

\section{Background and Related Work}
\label{sec:rw}

In this section, we review relevant literature on Well-Being Coaching (WBC) and other mental well-being practices, previous studies examining well-being applications in Human-Robot Interaction (HRI), as well as the use of \textcolor{black}{User-Centred Design (UCD)} and Qualitative Research in HRI for well-being.

\subsection{Well-being coaching and other mental well-being practices}

Well-being coaching is a non-clinical practice which focuses on improving the coachee’s mental well-being, goal-striving and hope \cite{green2006cognitive}. Coaching (cf. psychological therapy aimed to address mental illness) emphasizes the present, as well as how the coachee may flourish in the future, and maximize their fulfilment in life and work \cite{hart2001coaching}. The usual structure of coaching in a one-to-one session is to clarify the coachee’s issues, establish clear, measurable and attainable goals, develop action plans, and review progress in the next session \cite{green2006cognitive, grant2014solution}. Issues addressed by coaching can range from lifestyle (e.g. smoking cessation), to working life (e.g. learning a new skill), to personal relationships (e.g. addressing an issue with a partner). The coach’s goal is to facilitate and activate the coachee’s thinking and progress toward the goal. Coaching can emphasize different psychological practices, e.g. cognitive behavioural (focus on the relationship between thoughts, feelings, and actions) and brief-solution focused practice (helping the coachee focus on how their problem is already being addressed in small ways) \cite{green2006cognitive}, or positive psychology (paying greater attention to the positive aspects of one's life) \cite{seligman2007coaching}. Another style of coaching is Life Coaching, where goals can include dealing with stress, creating more meaningful relationships, and creating a more fulfilling and purposeful life in general. Life coaches can use varying tools such as drawing, writing, body-awareness exercises or visualization \cite{grant2010life}.

Other forms of well-being practices that serve to improve the practitioner's mental well-being are mindfulness and meditation, which can improve the practitioner's well-being and reduce negative emotions \cite{eberth2012effects}. Mindfulness training is often integrated into a meditation practice, and such meditation has shown to e.g. decrease stress and anxiety in college students \cite{bamber2016mindfulness}. Mindfulness has experienced a rise in popularity recently, with it being applied to workplace environments \cite{johnson2020mindfulness} as well as seeing new digital applications to conduct mindfulness training \cite{mrazek2019future}.
An important difference between mindfulness training and well-being coaching, when considering from a digital implementation perspective, is that the first can be conceptualized as instructing the participant in a particular skill, while the second is a conversation-based practice that trains people to focus on small things in their lives. This distinction may lead to different design recommendations for robotic well-being coaches operating in this area. 
It should be noted that both forms of well-being practices have already been recreated and commercially available in mobile application form: the mindfulness meditation application Headspace \cite{howells2016putting} has been shown to improve depressive symptoms and improve resilience \cite{flett2019mobile}, and the application Woebot was applied to deliver cognitive-behavioural therapy to young adults and reduced depressive symptoms \cite{fitzpatrick2017delivering}.

\subsection{Mental Well-being in HRI}

While robots have been explored for clinical interventions (such as communication therapy for children with autism spectrum disorders (ASD) \cite{scassellati2012robots} and social commitment robots for elderly people with dementia \cite{mordoch2013use}), applications focusing on mental well-being for non-clinical populations \hl{of different ages} are very few.

For example, \citet{bodala2021teleoperated} examined the participants' perceptions of a robotic vs. a human mindfulness coach in a longitudinal study, finding that while the human coach was rated significantly higher than the robotic coach, both received positive feedback. The study targeted a non-clinical population and excluded participants with high anxiety or depression levels prior to the study. Additionally, participants' \emph{Neuroticism} \hl{(i.e., nervousness, moodiness, and temperamentality)} and \emph{Conscientiousness} \hl{(i.e., organization, thoroughness, and reliability)} personality traits \hl{(as defined in the OCEAN model of personality} \cite{norman1963toward, goldberg1993structure}) were found to influence their perceptions of the robot.  
Also, \citet{jeong2020robotic} conducted a longitudinal study with the robot Jibo enabling positive psychology interventions for students, showing improvements in participants' well-being, mood, and readiness to change. Findings from this study also confirmed that participants' personality traits of neuroticism and conscientiousness influenced the interventions' efficacy. This study also targeted a non-clinical population, although no selection method is mentioned. Robots have been also used to disclose feeling and thoughts. For instance, \citet{akiyoshi2021robot} explored how to make people’s moods more positive through conversations about their problems with a robot, \hl{as compared before and after the conversation with the robot}. Their findings showed that people who interacted with the robot  self-disclosed more and experienced less anger than those who did not use the robot. \citet{duan2021self} ran an empirical study to compare self-disclosing in a diary journal or to a social robot after negative mood induction, targeting a population of people with depression. Their results showed that people who felt strongly negative after the negative induction benefited the most from talking to the robot, rather than writing down their feelings in the journal.
Additionally, a preliminary study for a robot displaying mood data, applying methods of community-based discussions with students, was conducted to improve the data visualization methods of such a robot \cite{karim2022community}.
Arts-based theatre methods were used in a pilot study with a Nao robot to address well-being with older adults living in a residential care setting \cite{fields2021shall}. The 3-session study found that depression and loneliness scores decreased in the study population \hl{during the study, which involved interacting with a robot}.
A recent study also explored how a robot can aid in the evaluation of children's mental well-being \cite{abbasi2022can}. These results demonstrated that the mental well-being evaluation with the robot seems to be the most suitable in identifying well-being related anomalies as compared with the self-report and the parent-report standard tests.

\subsection{User-centred methods in HRI for well-being}

\emph{User-centred} HRI aims to incorporate users' perspectives and needs into the design of robotic applications \cite{kim2011user}. 

Participatory Design (PD) refers to actively involving stakeholders (e.g. prospective users and relevant subject domain experts such as well-being coaches) in the design process as active co-designers \cite{spinuzzi2005methodology}. 
User-centred design has been previously used to design applications aimed at mental well-being in HRI, mainly in clinical contexts. Some authors also noted creating design guidelines or recommendations for robots in their respective applications at the end of the design process. \citet{lee2017steps} employed PD to design robots for various social contexts together with older adults with depression, taking into account the participants' issues and concerns in a socially responsible way.  \citet{winkle2018social} used PD to design robots for rehabilitative therapy together with therapists in focus groups, and presented design implications for robots and best practices based on therapists' knowledge. 
 \citet{axelsson2019participatory} used PD to design a robot to teach sign language to children with autism. 
The authors presented design guidelines for future robots for children with ASD.  \citet{moharana2019robots} employed PD to design robots for dementia caregiving together with dementia caregiver support groups, broadening the context of dementia robot design to include informal family caregivers in the process. The authors introduced design guidelines for such robots, and elucidated how robot attributes should change according to the stage of dementia of the caregivee.

\hl{To the best of our knowledge, none of the past works have defined a set of design and ethical guidelines for robotic mental well-being coaches that can inform the field on how to design and develop them. This paper leveraged on the results of three user-centered studies involving human mental well-being coaches and coachees to distil this set of guidelines as described in the following sections.}

\section{Methodology}
\label{sec:methodology}

This section introduces the concepts of Thematic Analysis (TA), which we use to collate and analyze the data from the three studies \hl{conducted (described in detailed respectively in Section} \ref{sec:study1}, \ref{sec:study2} and \ref{sec:study3}), as well as \textcolor{black}{a qualitative analysis inspired by the} Qualitative Meta-Analysis, which we employ to present a cohesive whole out of these three studies, as well as what we base our design guidelines and recommendations on.

\subsection{Thematic Analysis}
\label{sec:ta}

Our three studies use Thematic Analysis (TA) as a method to analyze qualitative data collected from interviews and workshops. This method has been previously used to examine robotic well-being applications \cite{winkle2019mutual, moharana2019robots, cagiltay2020investigating}. For example, \citet{moharana2019robots} used grounded theory to organize data collected from interactions with family caregivers into 16 major themes encountered in dementia care. The researchers used the themes to create scenarios of dementia care, which were then used for robot design.
In each of the three presented studies, we employ the 6-step method exemplified by \citet{braun2006using}. The 6-steps consist of: 1) familiarizing ourselves with the data (i.e., transcribing the data, reading them, and noting some initial ideas), 2) creating initial codes (i.e., identifying the codes within the dataset and collating data to the corresponding code), 3) searching for themes (i.e., collating codes into themes collecting all data under the relevant theme), 4) reviewing the themes (i.e., checking if the themes identified work also in relation with the codes), 5) defining and naming the themes (i.e., generating specific definitions and names for each theme, consistently with the story of the whole dataset collected), and 6) finally creating a report (e.g., extrapolating examples for each theme). We applied a grounded theory approach --- defined as the analysis of generating a reasonable theory of the phenomena that is grounded in the data \cite{mcleod2001using} --- where the themes in all three studies were selected on the basis of what we found in the data.

\subsection{Qualitative Analysis}

We apply a grounded-theory \textcolor{black}{qualitative analysis inspired by} Qualitative Meta-Analysis on the results of these three studies, in order to synthesize a summary, and to examine themes in common, as well as themes that differ and why these differences may arise \cite{timulak2009meta, timulak2014qualitative}. \textcolor{black}{We conduct this multi-study analysis to provide a ``more comprehensive description'' \cite{timulak2009meta} of how robotic well-being coaches should be designed taking into account users' and professional coaches' perspectives. In our analysis, we take into account the different context of each study, i.e., they were conducted at different phases of the iterative design process. This focus on the analysis of consecutive and iterative design phases makes it distinct from traditional Qualitative Meta-Analysis.} To conduct this \textcolor{black}{qualitative analysis}, we examined the results of the TA conducted on all three studies, how themes were repeated through each study (congruence), and how the results of the studies differed from each other (divergence). We analyze how these convergences and divergences may have resulted due to either differences in opinion, or differences in the study structures. \textcolor{black}{We do this in order to both highlight the findings of the different studies, as well as how these findings were influenced by the different context in different phases of the iterative design process.} Based on this \textcolor{black}{qualitative analysis}, we propose design guidelines for the creation of future robotic well-being coaches. We argue that our three studies and the resulting \textcolor{black}{qualitative analysis} are sufficient to propose such guidelines, as we observe \emph{data saturation}, i.e. the point where no new information are observed within the data \cite{guest2006many}. Our original three studies address the design problem of robotic mental well-being coaches from different perspectives and with different participants, and we see similar themes repeated throughout all studies. Additionally, we pay attention to the differences in these studies, as these can present important contextual differences in specific mental well-being applications or participant populations.

\section{Iterative Design Process: Three User-centred Studies}

\hl{In order to design the robot coaches to promote mental well-being, we followed a three-step iterative design process consisted of three user-centred studies. Firstly, we held an exploratory \textcolor{black}{user-centred} design study (described in Section} \ref{sec:study1}) \hl{in which both well-being coaches and university students' prospective users of mental well-being practice who had no to very little experience with robots participated. Our results suggested that the robot's ability to adapt and recognize users' emotions were crucial for the success of the practice delivered by a robotic coach. This motivated our second user-centred study (described in Section} \ref{sec:study2}), \hl{in which participants (again University of Cambridge's students) interacted with a robotic coach featuring adaptive capacities to deliver positive psychology exercises. The collaboration with one of the coaches in Study 1, who had expertise in positive psychology, motivated us to focus on this particular coaching style. We then concluded our design process by conducting another \textcolor{black}{user-centred} design study which included the participants of Study 2 -- who had interacted with the robotic coach -- as well as the well-being coaches (see Section 
}\ref{sec:study3}) \hl{and focused on the evaluation of various applications of robotic coaches to promote mental well-being.}

\subsection{Study 1}
\label{sec:study1}

The first study aims to collect and examine the expectations and perceptions of prospective well-being users and well-being coaches. This study was originally discussed in \cite{axelsson2021participatory}. Here, we summarize the study, and present aspects related to well-being practices that were out of scope for the prior publication.   
\subsubsection{Participants}
We involved 11 participants, 8 of them were interested in well-being practices with a robotic coach \hl{(as reported by themselves when asked by the researchers)}, and 3 of them \hl{(2 females and 1 male)} were well-being coaching professionals, specialized in Solution-Focused Practice (SFP), Mindfulness, and Life Coaching respectively. We conducted group discussions with the 8 prospective users, and semi-structured interviews with all participants, including the 3 well-being coaches. 
The prospective users had taken part in either SFP sessions with a human coach \hl{(6 participants, details of study and practice reported in }\cite{axelsson2021participatory, cheong2023}), or Mindfulness sessions with \hl{a human coach} \hl{(2 participants, details of study and practice reported in }\cite{bodala2021teleoperated}). \hl{In the SFP and Mindfulness studies, participants interacted with a coach conducting Mindfulness sessions (5 sessions, in a group) or SFP sessions (4 sessions, individually). Participants in both studies }were recruited from the University of Cambridge student and postdoctoral researcher population. \textcolor{black}{Further demographic details of the participants were not collected in this study.}

\subsubsection{Setting}
\begin{figure}[h!]
  \centering
  \includegraphics[width = 0.65\textwidth]{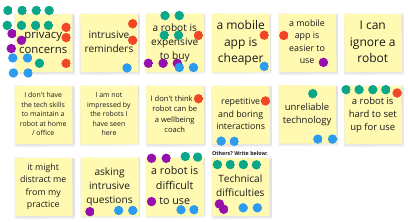}
  \caption{Results of participant votes using the Miro board to indicate what would stop them from using the robotic well-being coach. Originally published in \cite{axelsson2021participatory}.}
 
  \label{fig:disadvantages_study1}
\end{figure}

Due to COVID-19 restrictions, we performed both the group discussion with the prospective users and the semi-structured interviews online (via MS Teams).  
To facilitate the group discussions in the \textcolor{black}{user-centred} session online, we asked participants to use the online tool \textit{Miro}\footnote{https://miro.com} (an example of working with Miro can be seen in Figure \ref{fig:disadvantages_study1}).

\begin{table}[h!]
    \centering
    \begin{tabular}{ll}
    \toprule
        \textbf{Items} &  \textbf{Duration}\\
        \midrule
         Pre-discussion survey (in writing) & 5 min \\
         \midrule
         Introduction & 3 min\\
         Warm-up discussion about well-being practices & 10 min \\
         Introduction to social robots and demo videos & 7 min \\
         Ideating robotic well-being coach & 15 min \\
         Discussion on robotic well-being coach features and capabilities & 20 min \\
         Conclusion & 2 min \\
         \midrule
         Post-discussion survey (in writing) & 5 min \\
         \bottomrule
    \end{tabular}
    \caption{Structure of the 
    group discussions. Individual 30-45 minute interviews performed beforehand. Originally presented in \cite{axelsson2021participatory}.}
    \label{table:study1-structure}
\end{table}

\subsubsection{Protocol}
This study focused on the prospective users' and coaches' experiences when participating in and coaching well-being practices, and how technology has helped or hindered these tasks. It consisted of group discussions, semi-structured interviews, and a canvas filling (optional) \cite{axelsson2021participatory}. 

We first conducted 8 one-to-one semi-structured interviews with the prospective users for 30-45 minutes, to ask them about their experiences with well-being practices, well-being practices involving technology, and potential improvements and challenges they had encountered (see Appendix \ref{app:study1-user-interview} for an example interview structure). Additionally, we interviewed the 3 well-being coaches about their experiences with different coaching practices, how they had previously used technology, and how they could see a robot conducting coaching (see Appendix \ref{app:study1-coach-interview} for an example interview structure).

After the interviews, we organized group discussions with the same 8 prospective users we had interviewed. The discussion topics were informed by the interviews conducted with the prospective users and coaches. We divided the group discussion participants into two groups (4 participants each), and each group discussion followed the same structure (see Table \ref{table:study1-structure}). Each group discussed their previous experiences with well-being practices, how they could see a robot coaching well-being practices, and what capabilities and features such a robotic coach should have.

The full list of topics discussed during the group discussion was as follows: \hl{(1) Well-being practices (e.g., Describe with one word your experience participating in the SFP / Mindfulness study), (2) Aim of the group discussion (e.g., Whether and how social robots can used as coaches for well-being), (3) Ideation on a robotic well-being coach (e.g., How do you think social robots could provide or support well-being practices, or work as a well-being coach?), (4) What should the robot be like? (e.g., What do you want the robot to do / say?). Please find more detail of the study in} \cite{axelsson2021participatory}.

Three of the prospective users as well as two well-being coaches also participated in an optional part of the study, where they filled in a subset of the Social Robot Co-Design Canvases \cite{axelsson2021social, Axelsson_2021}, to further detail design considerations in a robotic well-being coach.

\begin{figure}[h!]
  \centering
  \includegraphics[width = 0.65\textwidth]{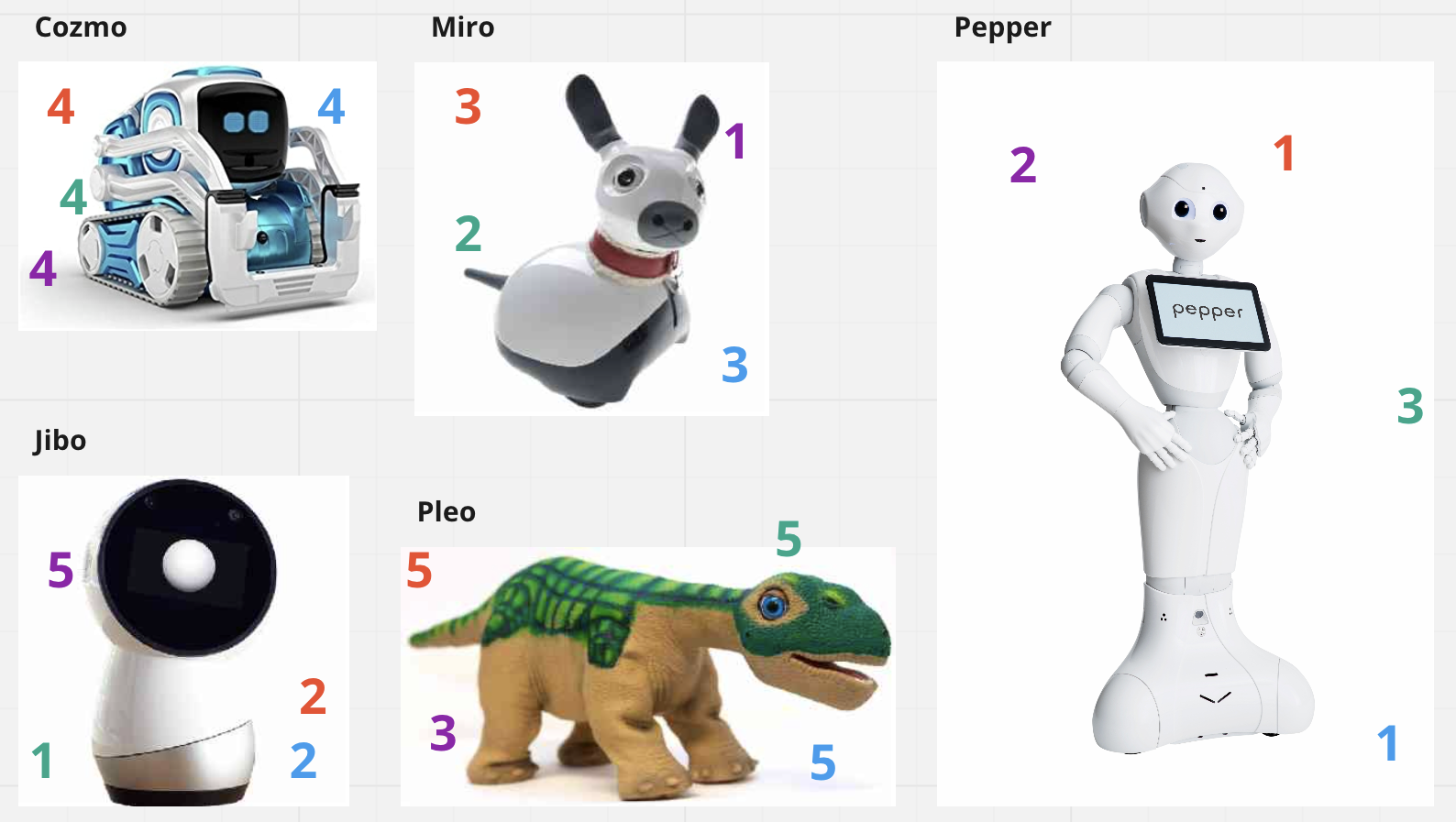}
  \caption{Results of participants' ranking robot embodiment for well-being coaching (discussion group of four participants, each having their own colour). 
  While Pepper was preferred by this group, Miro and Jibo were also preferred by some, indicating that preferences vary across individuals. Originally published in \cite{axelsson2021participatory}.}
  
  \label{fig:voting}
\end{figure}

\subsubsection{Data analysis}
A Thematic Analysis (TA) was conducted, following the procedure described in Section \ref{sec:ta}, based on data transcribed from interviews (both with prospective users and coaches) and group discussions, as well as data collected from the Miro board and the Social Robot Co-Design Canvases.

\subsubsection{Summary of Main Findings}

\begin{figure*}[t!]
  \centering
  \includegraphics[width=\textwidth]{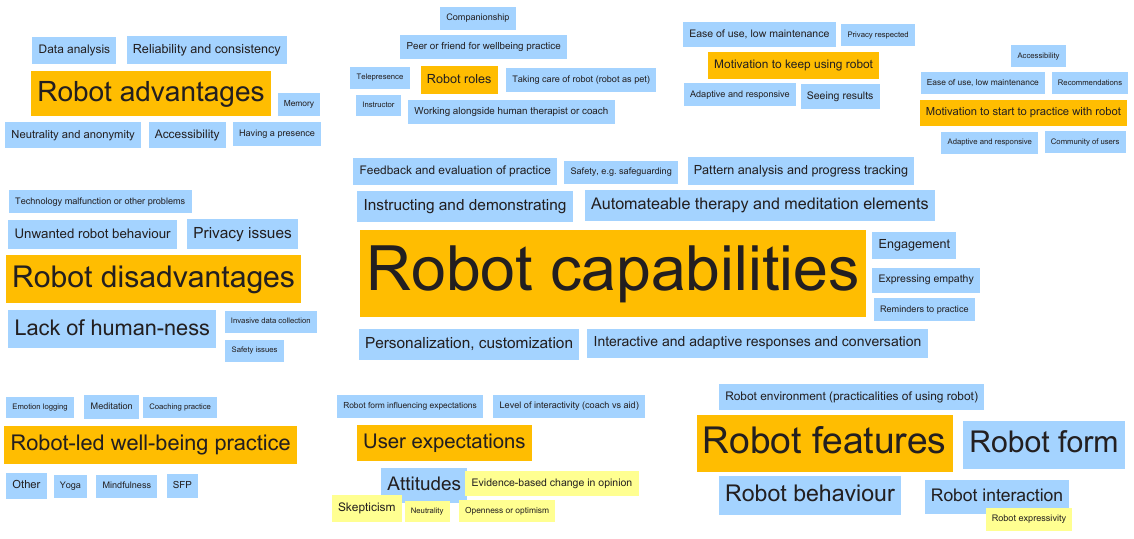}
  \caption{Study 1: Themes defined in the TA are presented in orange, while codes related to these themes are presented in blue, and sub-codes are presented in yellow (best viewed in colour), from \cite{axelsson2021participatory}.}
  \label{fig:thematic1}
\end{figure*}

The TA resulted in 9 major themes (as seen in Figure \ref{fig:thematic1}). \hl{This section summarised the main findings of the study}. Quotes on the topics summarized in this paper as well as images from the group discussions \hl{and more detailed and extensive description of the results} can be found in \cite{axelsson2021participatory}. 
In the following sections, we refer to the prospective users with the term "participants" while we specify which quote has been told by the coaches directly naming them.

\paragraph{Well-being practices}

Both prospective users and coaches were asked to think about what \emph{well-being practices} they could see a robot leading, bringing up \emph{mindfulness, meditation, yoga} and \emph{SFP}. Participants detailed what kind of \emph{role} the robot should take during such practices (e.g. a \emph{instructor}, a \emph{peer or friend} for the practice, or a \emph{supplement for instructions from a human coach}). Participants perceived that the robot should have differing roles depending on the practice the robot would be conducting, e.g., for meditation the robot could act as more of a peer, and for SFP more as an instructor. The practices discussed were drawn mainly from the participants' own previous experiences with well-being practices, either from their own practice or from the studies they had participated in. Comments made by the three professional coaches about how a robotic coach might conduct the practices they were specialized in can be seen in Table \ref{table:coachWBPstudy1}, e.g., examining the specific types of SFP techniques (The Miracle Question) and Meditations (Mindfulness of Breathing and Loving Kindness) they could see the robot having an advantage in.

\paragraph{Robot capabilities}
\label{sec:study1rob_cap}

A major focus during group discussion and coach interviews was on \emph{robot capabilities} during well-being coaching, i.e., what capabilities participants thought that the robot should display during its interactions with coachees, in order for them to benefit from the interaction. Participants thought that a RC could \emph{engage participants in the practice} by providing a \emph{sense of presence}, or by \emph{doing the practice together with them}. They noted that the physical presence of the robot could help them better focus on the practice \hl{(in comparison to e.g. a mobile app, which has no presence)}. Participants also noted that they would like the robot to \emph{express empathy} by \emph{reacting to how they feel} and to \emph{praise} them in suitable situations. This would help them feel that the robot was acknowledging their emotions. Participants also hoped that the robot would display empathy by changing its \emph{expressions}, however the SFP coach remarked that it would be important to have the correct level of expressivity in order to mitigate the \emph{uncanny valley} effect that could arise due to incongruent responses.

Robot \emph{responsiveness} in terms of \emph{interactive and adaptive responses} was also discussed. Participants said they would like the robot to \emph{summarize their thoughts} to give a sense of \emph{acknowledgement}. A participant noted that such a sense could also be achieved through \emph{expressive behaviours and movement} instead of tailoring verbal utterances.

\paragraph{Robot features}

Robot features, i.e., how the robot should look, interact, behave, and how its environment should be shaped --- were discussed at length. Participants discussed the robot's \emph{form}, i.e. its appearance and embodiment. Participants were asked to consider 5 robots (Jibo, Pepper, Miro, Cozmo and Pleo) and rank them as most appropriate to least appropriate as a RC. Jibo received the highest points, followed closely by Pepper and Miro. Participants had differing opinions on whether they would prefer a humanoid or an abstract robot, but the consensus was that the robot's function should match its form. For example, participants noted that they would not expect the robotic dog Miro to talk, and would expect quite sophisticated, human-level \emph{interaction} from a humanoid robot such as Pepper. Participants expressed that a two-way communication with a humanoid robot should be designed carefully, so that the robot is seen to understand and respond appropriately, and to avoid incongruous responses.

Robot \emph{behaviour} was discussed from the perspective of being too \emph{forceful} or \emph{patronizing} vs. \emph{not direct enough}. Participants wanted the robot to be encouraging, but they noted a balance is needed for this not to be patronizing. Participants explained that they want feedback on their practice as well as reminders, but these should be given on the appropriate level, and not be too intrusive. \emph{Responsiveness} was also seen as important, as discussed in length as part of the robot's capabilities.

\paragraph{Robot advantages and disadvantages}

Coaches and participants shared what they perceived to be the potential advantages and disadvantages of a RC, especially in comparison with a human coach, or a hypothetical mobile application with a well-being coaching function. While many participants noted that a robot could not replace a human as a primary method of delivery for well-being practices, it was seen to have the advantage of \emph{accessibility} in comparison to a human coach (both in terms of timing and affordability). A robot's \emph{physical presence} was seen to be a major advantage over a well-being practice being administered via mobile app or other digital means. However, \emph{technological problems} such as poor internet connection or lack of software updates were seen to impede the potential advantage of accessibility.

Participants noted that the \emph{lack of humanness} of a RC could be a major disadvantage. For instance, \emph{unwanted robot behaviour}, such as incongruent responses when misunderstanding the user, could lead to participants feeling less connected or the session getting sidetracked. However, a RC could provide \emph{reliability, consistency and uniformity} that a human coach could not. A coach noted that a robot would not get tired and could thus accommodate more participants, and would be more \emph{consistent} and maintain \emph{uniform} interactions due to not being affected by a daily routine, personal life, or a coach's personal factors such as nervousness or maintaining sufficient emotional distance. 

\emph{Analyzing feedback data} was seen as a potential advantage by the mindfulness coach. Participants also noted that a robot could record the practice, and adapt the practice according to data gathered. However, a disadvantage related to this was \emph{invasive data collection}, where the SFP coach specifically mentioned that too much data such as biological signals could affect outcomes by making participants even more anxious. Participants were also concerned about \emph{privacy issues}, although the SFP and life coach noted that a robot may actually provide more privacy if data is appropriately protected, since there is no room for human error. 

\emph{Neutrality and anonymity} was the final advantage of a potential RC perceived by the coaches. The SFP coach noted that a robot could be neutral in gender, and less intimidating than a human coach if it were smaller in size. The robot could also be \emph{less judgemental}, as noted by the SFP coach and meditation coach. The life coach also noted that a robot could be more neutral due to no bias through its \emph{lack of life history}.

\begin{table*}[t]
\small
\centering
  \caption{Quotes from coaches ($Ci, (i=1,2,3)$) regarding suitable well-being practices for a robotic well-being coach.}
  \label{table:coachWBPstudy1}
  {
  \begin{tabular}{|C|L|}\Xhline{1.0 pt}
    \rowcolor{gray!40}
    \textit{\textbf{Well-being Practices}} & \textit{\textbf{Quotes from coaches}}\\ \Xhline{1.0 pt}
    \rowcolor{gray!15}
    Solution-Focused Practice  
    & {\textbf{The Miracle Question}
    
    C1: ``The miracle question. You go to sleep and a miracle happens in the night. [...] The reason for the problem disappears and it solved the problem [...] but you were asleep and you didn't know the miracle that happened. What would be the first signs that something had changed, [and you] amplify what [the participant] said. [...] They’ll say, ‘This thing would happen’ and [you ask] ‘What else, what else? [...] So there are some questions that are very standard.''
    
    C1: ``[...] the words are very carefully chosen for a certain effect, so that could be very easily automated. It’s already scripted basically [...].''}   \\
    \Xhline{0.5 pt}
    \rowcolor{gray!40}
    Meditation / Mindfulness     
    & {\textbf{Mindfulness of Breathing}
    
   C2: ``[...] it's four stages. [...] You start with counting your breaths, after you breathe out, then breathe in. Then you stop, drop, [...] counting. You concentrate on the touch of the breath. Feel it. It's easy to instruct.'' 
   
   C1: ``[...] Breathing exercises (not just mindfulness or safe place visualisation) would be a good skill for a robot to be able to teach. It would have good pacing and could talk through the breathing script whilst also seeming to practise with you (without running out of breath which is hard as a human!).''
   
   \textbf{Loving Kindness (Metta Bhavana)}
   
   C2: ``It's a bit more complicated [than Mindfulness of Breathing] as you need to [...] give [the participants] a bit more complex [information]. I'm sure you can do that because if a recording can do that, a robot might do that as well.''}   \\
    \Xhline{0.5 pt}
    \rowcolor{gray!15}
    Life Coaching
    & {\textbf{Mindfulness Exercises}
    
    C3: ``These exercises [are already] recorded, audio with some relaxation exercises or imagination exercises, [and] guiding people through those.''

    \textbf{Other Exercises}
    
    C3: ``[...] Many of these exercises, whether they are [...] drawing, writing, answering questions. They're ready exercises so they can be recorded or they can be documented. And then the robot could easily show these in the written format or talk [to the participants].''
    
    C3: ``[...] The Pepper robot had the screen there, so if it  [were a] touch pad type of screen, a person could draw there, write there, [...] at least some words. There could be, in the same area where they meet the robot, [...] paper where they could write.''}
    \\
    \Xhline{0.5 pt}
    \rowcolor{gray!40}
    Cognitive Behavioural Coaching
    & {\textbf{Fear Hierarchy}
    
    C1: ``CBT is quite like you’re teaching certain skills, like graded exposure might be part of CBT [...].'' 
    
    C1: ``[...] Developing a fear hierarchy and helping someone. And track their progress. But I think you'd want it to be in tandem with a person.''
    
    \textbf{Behavioural Activation}
    
    C1: ``This when in particular people are depressed. You can do behavioral activation, which is a part of CBT, which is scheduling in activities that are meaningful for someone and tracking how that makes them feel.  I could see that kind of activity scheduling working in a robotics context.'' 
    
    C1: ``But I think you would want it alongside a person who helped you generate ideas.''}   \\
    
    \Xhline{1.0 pt}
  \end{tabular}
  }
\vspace{-5mm}
\end{table*}

\newpage

\subsection{Study 2}

\label{sec:study2}

\hl{Our findings in Study 1 showed that the adaptability and emotional responses are important aspects for a robotic coach to deliver mental well-being practices. With the help of one of the mental well-being coaches who participated in Study 1 and motivated by those results, we} designed the behaviour and interactions of a Robotic Coach (RC) to perform a brief, one-off positive psychology session with healthy adults as a proof-of-concept. \hl{We focused on the positive psychology practice given the expertise of the human coach who collaborated with us for designing this study. }
We aimed to examine the \emph{robot capability} (see Section \ref{sec:study1rob_cap}) of \emph{responsiveness and adaptation}, specifically displaying \emph{empathy} via the robot's \emph{emotional expressions}, in response to participants sharing their experiences. This study is originally presented in \cite{churamani2022continual}\hl{, where we applied Continual Personalization to accomplish the adaptation capability.}

\subsubsection{Participants}
In a one-session human-robot interaction study, 20 participants \hl{(12 females, 5 males,
3 not disclosed, aged 26.70} $\pm$ \hl{3.68 years old from 12 different nationalities)} recruited from the University of Cambridge student population interacted with the RC Pepper conducting Positive Psychology exercises  over a $30 \pm 11$ minute session. 
The study consisted of three interaction rounds, each containing three exercises: 1) Two Impactful Things, 2) Two Things You Are Grateful For, and 3) Two Accomplishments. Each interaction round focused on the past, present, and future, respectively. The study examined the use of affect-based adaptation using Continual Personalization. The implementation of this adaptation is out of scope for this paper, and is described in more detail in a paper detailing the technical implementation of the robot and the corresponding interaction flow \cite{churamani2022continual}.

\subsubsection{Setting}

\begin{figure}[t]  
    \centering
    \includegraphics[width=0.45\textwidth]{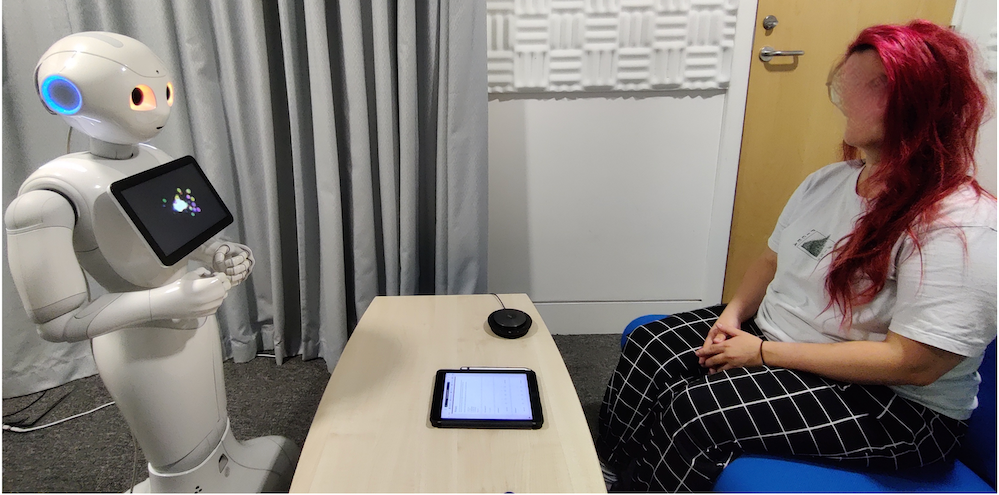}
    \caption{Participant and Pepper interacting with each other.}
    \label{fig:exp2_setup} 
\end{figure}

The study was conducted in a dedicated room at the University of Cambridge. It was equipped with two cameras facing the robot and the participant, in order to capture the interaction. The Pepper robot's on board RGB camera was used to capture the participants' facial affect. An external microphone was used to capture the participants speech, and Pepper used its on board speakers to communicate with participants.
The participants were sat in front of the Pepper robot, with a table separating them in order to create a natural distance, and with the tablet containing the questionnaires to be completed during the study placed on the table (see Figure \ref{fig:exp2_setup}).

\subsubsection{Protocol}

\begin{table}[h!]
    \centering
    \begin{tabular}{ll}
    \toprule
        \textbf{Items} &  \textbf{Duration}\\
        \toprule
         Pre-interaction survey (writing) & 5 min \\
         \midrule
         Robot introduction & 3 min\\
         Explanation of PP & 3 min \\
         ``Past'' round & 10 min \\
         ``Present'' round & 10 min \\
         ``Future'' round & 10 min \\
         Conclusion & 3 min \\
         \midrule
         Post-interaction interview & 8 min \\
         \bottomrule
    \end{tabular}
    \caption{Structure of the study 2. Average times for each session were: past $(11 \pm 4 minutes)$, present $(9 \pm 3 minutes)$, future $(10 \pm 3 minutes)$, and overall interactions $(30 \pm 11 minutes)$. Originally presented in \cite{churamani2022continual}.}
    \label{table:study2-structure}
\end{table}

The structure of the one-time interaction sessions can be seen in Table \ref{table:study2-structure}. Each interaction session consisted of a pre-interaction survey, introduction phase (with explanation of PP),  three rounds, focusing on the participants' life in the \emph{past}, \emph{present}, and \emph{future}, and a final post-interaction interview. The HRI script design was based on pre-existing positive psychology exercise literature, and was examined with a professional coach before the study, in order to mitigate any potential negative perceptions in the participants.

Before starting the interaction (pre-interaction survey), participants were asked to fill out few questionnaires that are described in details in Section \ref{sec:st2survey}.

Each interaction round (\emph{past}, \emph{present}, \emph{future}) consisted of three exercises. Each participant conducted all three exercises three times, once for each interaction round.
\begin{enumerate}
    \item[i.] \textit{Two Impactful Things:} In this warm-up exercise, the robotic coach asked participants to talk about two impactful things or events that either happened in the past two weeks (\textit{past}), happened or are expected to happen today (\textit{present}), or are expected to happen in the coming two weeks (\textit{future}). The robotic coach also asked the participants to think about why these events happened and how they made them feel. 
    \item[ii.] \textit{Two Things the Participant is Grateful For:} This exercise focused on developing \emph{gratitude}; a concept emphasised during Positive Psychology. Focusing on gratitude can increase positive affect, subjective happiness and life satisfaction~\cite{cunha2019positive, emmons2002gratitude}. The robotic coach asked the participant to recall or imagine (depending on \textit{past, present} or \textit{future} interactions) two things that they felt or might feel grateful for.

    \item[iii.] \textit{Two Accomplishments:} In this exercise, the robotic coach asked the participant to think about \textit{past, present} or \textit{future} \emph{accomplishments}, focusing on self-esteem, which has been applied to increase well-being and ameliorate depressive symptoms~\cite{gander2016positive}. The robotic coach asked the participant to describe past, present or potential accomplishments, strengths the participant applied or may apply to accomplish these~\cite{lopez2018positive}, and how these accomplishments make the participant feel. 

\end{enumerate}

After finishing all three interactions with the robotic coach (post-interaction interview), a semi-structured interview was conducted with the participants, documenting their perceptions of the robot and the positive psychology exercises. The interview structure can be seen in Appendix \ref{app:study2-interview}.

\subsubsection{Surveys}
\label{sec:st2survey}

Prior to the interaction, participants filled surveys on a tablet (the IPIP-20 personality inventory, and the PANAS mood inventory). During the interaction, the participants filled the PANAS mood inventory after each interaction round (the past, present, and future), as well as the Godspeed and RoSAS questionnaires regarding their impressions of the robot. Participants also rated the statements ``The robot understood what I said'', ``The robot understood how I felt'', and ``The robot adapted to what I said and did'' after each interaction round on a 5-point Likert scale. 

\subsubsection{Data analysis}
The data collated and analysed in the next section were collected through the semi-structured interviews with the 20 participants conducted immediately after each interaction session. These interviews were analyzed with TA, the results of which are introduced in the next section. 

\subsubsection{Summary of Main Findings}

\begin{figure}[h!]
    \centering
    \includegraphics[width=\textwidth]{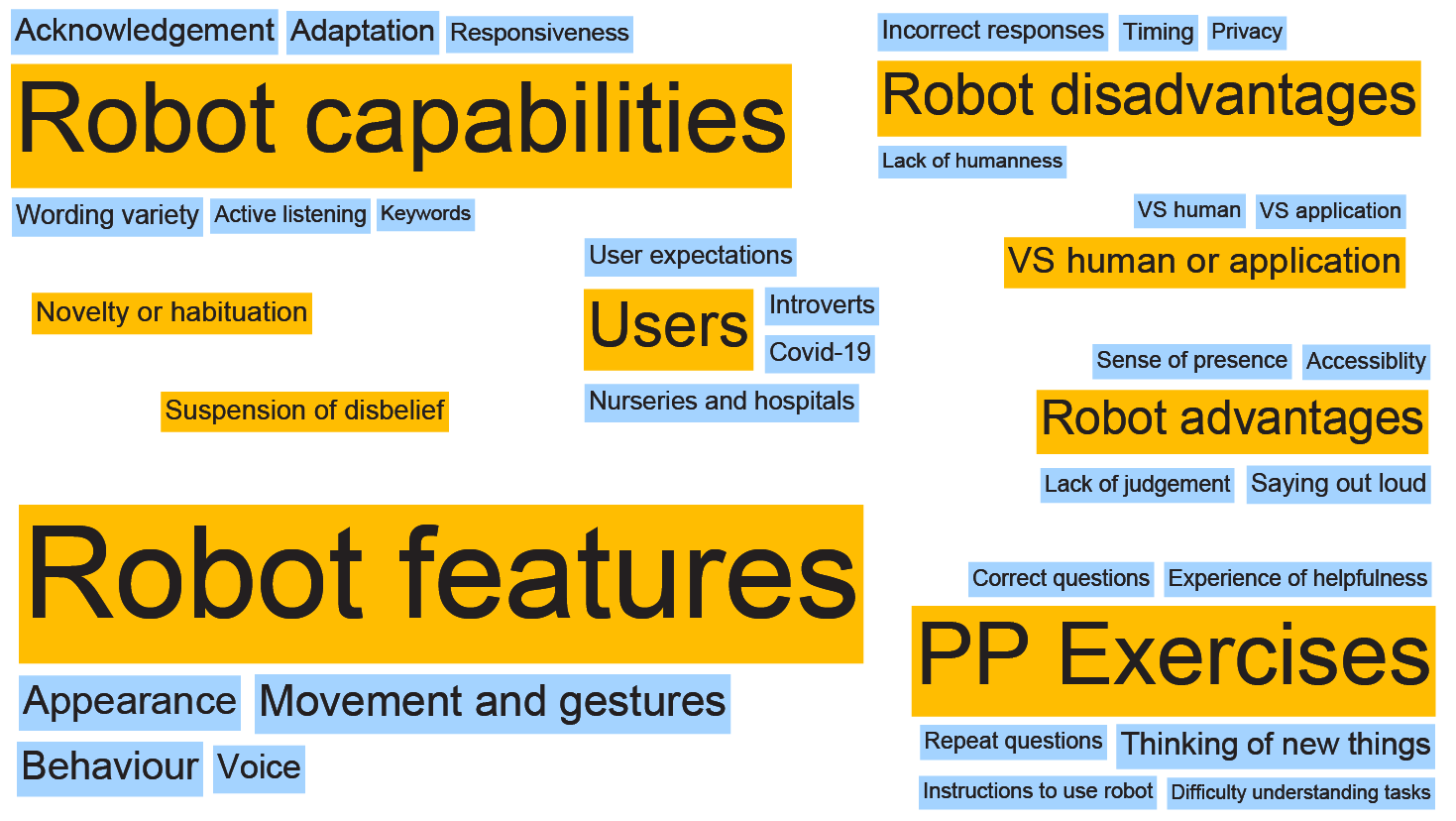}
    \caption{Study 2: Themes defined in the TA are presented in orange, while codes related to these themes are presented in blue (best viewed in colour).}
    \label{fig:thematic2}
\end{figure}

Participants were asked about their opinions on the performance of Pepper \footnote{https://www.softbankrobotics.com/emea/en/pepper} as a Positive Psychology well-being coach, what they thought worked well, and what they would change. The TA resulted in 10 major themes, with focus on robot features, capabilities, as well as the content of the Positive Psychology practices (as seen in Figure \ref{fig:thematic2}). A summary of the TA quotes can be seen in Table \ref{table:study2-quotes} and the complete list of results can be found in \cite{churamani2022continual}.

\begin{table*}[ht]
\small
\centering
\caption{Selected quotes from participants about study themes gathered during the semi-structured interview lasting for $\approx10$ minutes. The quotes have been slightly edited for clarity.}
\vspace{-1mm}
\begin{center}
\begin{tabular}{|C|L|}
\Xhline{0.5 pt}
\rowcolor{gray!40}
\textit{\textbf{Quotes about}} &
\textit{\textbf{Quotes from participants}}\\
\Xhline{0.5 pt}

\rowcolor{gray!15}
Novelty and Habituation &

{\textbf{P17}: ``I felt the last round [...] I got more used to the pattern [...], but it might have also been that it understood things more clearly.’’

\textbf{P11}: ``[...] it was weird to be talking to a robot. It's also the surprise when it first talked and moved, and then the first time it would say certain things that were kind of funny as well. It would say ‘oh thank you so much for sharing that with me’. [...] I think once I got used to it, and it didn't take very long to get used to it, the weirdness wore off early.'' 
}

\\
\Xhline{0.5 pt}

\rowcolor{gray!30}

Responsiveness and Adaptation &
{\textbf{P5}: ``There were a few times when it did have a specific new response or it said ‘That sounds like a great accomplishment’ or ‘You should be proud’ and that was nice, I was like ‘Oh thank you robot for listening to me’.’’

\textbf{P13}: ``[It had] good questions, but I don't think it adapted to anything I said [...] [It was] asking me questions that made me think productively.
[...] [But] it felt more like the robot has a good script than it was engaging in active listening.’’}

\\
\Xhline{0.5 pt}

\rowcolor{gray!15}
Exercises &
{\textbf{P7}: ``At the end I think I felt more grateful. [...] I'm feeling more positive than I was before the start, 
in that respect it was effective.’’

\textbf{P2}: ``It was good that it explained the thought process behind the exercises, explaining why thinking of positive aspects was important. [...] I guess [it’s] an exercise we could do ourselves, but it felt more interactive because the robot was asking you questions.’’

}\\
\Xhline{0.5 pt}

\end{tabular}
\label{table:study2-quotes}
\end{center}
\vspace{-6mm}
\end{table*}

\paragraph{Robot capabilities}

While participants enjoyed the content of the PP exercises, they would have preferred the robot to be more \emph{responsive}. Participants noted that when they were disclosing their feelings (during task 1, Two Impactful Things), the robot could have ``engaged more positively with their positive feelings''. \emph{Acknowledgement} of what the participants were saying, as well as a feeling of \emph{actively listening} could be done via movement, as well as adding \emph{wording variety} and having the robot respond differently to what the participants were saying (as opposed to repeating the same response every time). \emph{Adaptation} was requested from the participants, specifically the robot having the ability to ask follow-up questions and give suggestions based on the participants' responses. Participants said the robot could pick up \emph{keywords} from their speech, and adapt its utterances accordingly.

\paragraph{Robot features}

Participants had mixed impressions of the robot's \emph{behaviour}. Some perceived it as \emph{supportive}, \emph{compassionate}, \emph{friendly} and as \emph{giving space to talk} as well as \emph{positive reassurance}, and that they felt \emph{comfortable} and \emph{heard} during the interactions. However, others noted that it was \emph{robotic}, and even \emph{condescending} when telling the participants ``well done''. One participant noted being stressed out when talking to the robot, since they anticipated a quick response from the robot, and wanted to think of a response quickly so that the robot would not move on prematurely.

The robot's \emph{voice} was generally perceived well, with the \emph{slow pace} being called patient and listening. A few participants found the \emph{pronunciation} strange, with one noting being creeped out by it. However, another participant noted being positively surprised in that the voice wasn't as mechanical as they had thought.

The robot's \emph{appearance} was generally regarded positively. It was called \emph{cute}, \emph{friendly}, and even \emph{childlike and innocent}. The lack of facial expressions was perceived as a positive, with participants describing it as \emph{approachable} and ``taking in what I'm saying''. The human form factor was perceived as ``nice'', however being at ``the high end of humanoid before being too creepy''.

\paragraph{Robot advantages and disadvantages}

Participants made favourable comments on the robot in comparison to a hypothetical mobile app designed for the same set of exercises, noting that \emph{speaking out loud} was in itself helpful, as well as the \emph{sense of presence} that can be provided by its physicality. Advantages in comparison to a hypothetical human coach performing the same set of exercises were that the robot had a \emph{lack of judgement}, where it might be easier to talk to a robot due to it not being as reactive as a human, as well as \emph{accessibility}, where a human coach is limited to working hours and a robot can serve more people in a day.

However, a robotic coach was also perceived to have disadvantages. Participants noted that the robot had problems with \emph{timing}, where it sometimes interrupted participants before they had finished talking, and paused for too long between responses. These were perceived to be a distraction, and left participants feeling like they were being hurried up. The robot would also give \emph{incorrect responses} on occasion, robot's emotion recognition system made incorrect estimations of the emotion expressed by the participant. This made participants feel misunderstood. Another disadvantage of the robot was perceived to be \emph{privacy issues}, where participants were sensitive being recorded, and chose what to talk about on that basis. Finally, participants noted that while it could be an advantage in some cases, the \emph{lack of humanness} of the robot could be limiting in the depth of information it can ask, and couldn't give organic responses.

\paragraph{Positive Psychology exercises}

Participants especially made several comments on the Positive Psychology exercises, noting that they were \emph{helpful} and made them reflect, particularly about a positive future. They noted feeling more grateful and positive after the exercises, and that the robot was useful for guiding them through the exercises. Particularly, the robot asked the \emph{correct questions}, and that it explained the thought process behind the exercises. Participans noted that they \emph{thought of new things}, e.g. why they felt grateful, and what they felt grateful for. Participants noted that the robot could be improved by having the ability to \emph{repeat questions}. Some participants didn't hear or understand the robot the first time it explained a task, and wanted to be able to go back and ask for clarification. Some participants also had \emph{difficulty understanding the tasks}, noting that they weren't sure of the exercise structure. Here, too, it would have been helpful to be able to ask the robot for clarification. Some participants also noted that they wanted more \emph{instructions on how to use the robot}, e.g. how loud they should speak and how long they should wait before responding.

\paragraph{Users}

Participants talked about different aspects related to the \emph{users} of the robot. Specifically, they noted that the \emph{expectations} of the users coming to interact with the robot would shape the interactions. One participant noted that people who had not done therapy before, and thus did not have any expectations, could benefit from the robot the most. Another noted that it works on the basic level, but if a person had a bad day and was expecting more support, the robot could be a disappointment. 
One participant noted that the robot might be especially useful in the aftermath of \emph{COVID-19}, and that it could be useful for people isolated in homes due to the pandemic, or even people who are just socially isolated but would want a channel to open themselves back up to society. Other participants concurred with this, noting that the robotic coach could be particularly helpful for \emph{introverts} that felt shy or awkward with humans, and could feel more confident with a robot. Finally, one participant noted that the robot could be useful in \emph{elderly care homes and hospitals}, where the small interactions with a robot could make a huge difference where nurses weren't available, especially for the older generations who are alone.

\paragraph{Phenomena}
\label{sec:study2_phenomena}

Two Human-Robot Interaction related \emph{phenomena} were described by the participants. They are two separate phenomena, but are grouped here under the label \emph{Phenomena} since they were separate from other themes selected in the TA. 
\emph{Novelty and habituation} effects seemed to have an effect on participants' perceptions. In relation to the exercises themselves, one participant described understanding the exercises more clearly in the last round, as they had gotten used to them. However, another noted that they came to predict what would happen after the first round, and weren't reflecting as much anymore after that. Participants also described evolving perceptions of the robot, e.g. being initially surprised by it and laughing. Another described being initially creeped out by the robot's smile and some arm movements, but easy to get used to.
\emph{Suspension of disbelief} was the other HRI related phenomenon identified in the TA. Suspension of disbelief is a concept discussed in HRI \cite{duffy2012suspension} as the person's willingness to suspend their disbelief of what is living in physical social robotics, and how this willingess may affect people's perceptions of such social robots. The concept originally relates to an audience's willing participation in a fiction \cite{packman1991incredible}. Participants did not explicitly mention suspension of disbelief by name, however, many participants talked about feelings of strangeness and being conscious of the experience.

\newpage
\subsection{Study 3}

\label{sec:study3}
As this study is not presented in a prior paper, we present it here in more detail than Study 1 (Section \ref{sec:study1}) and Study 2 (Section \ref{sec:study2}).

\hl{The last step of the iterative design process consisted of another \textcolor{black}{user-centred} design study (Study 3) involving participants who interacted with the robot in Study 2.}
Study 3 was a group discussion aimed at examining what would motivate participants to use a robotic well-being coach in the long term, as well as what they would like to see from a robotic coach of Solution-Focused Practice (SFP). 

\subsubsection{Participants}
We recruited 2 well-being coaches \hl{(one of them had already taken part to Study 1 described in Section} \ref{sec:study1}), specialized in Solution-Focused Practice and Mindfulness respectively, as well as 3 prospective users who had \hl{tangible experience with the robot due to having} attended the Positive Psychology exercises conducted by Pepper (as presented in Section \ref{sec:study2} and in \cite{churamani2022continual}). \hl{Participants were 3 females and 2 males aged 32.8 }$\pm$ \hl{9.1 from 4 nationalities (British, Brazilian, Portuguese and Romanian), and 2 of them have some experience with the humanoid robots, while the other 3 had no to little experience with them.}

\subsubsection{Setting}

\begin{figure}[h!]
    \centering
    \includegraphics[width=0.5\textwidth]{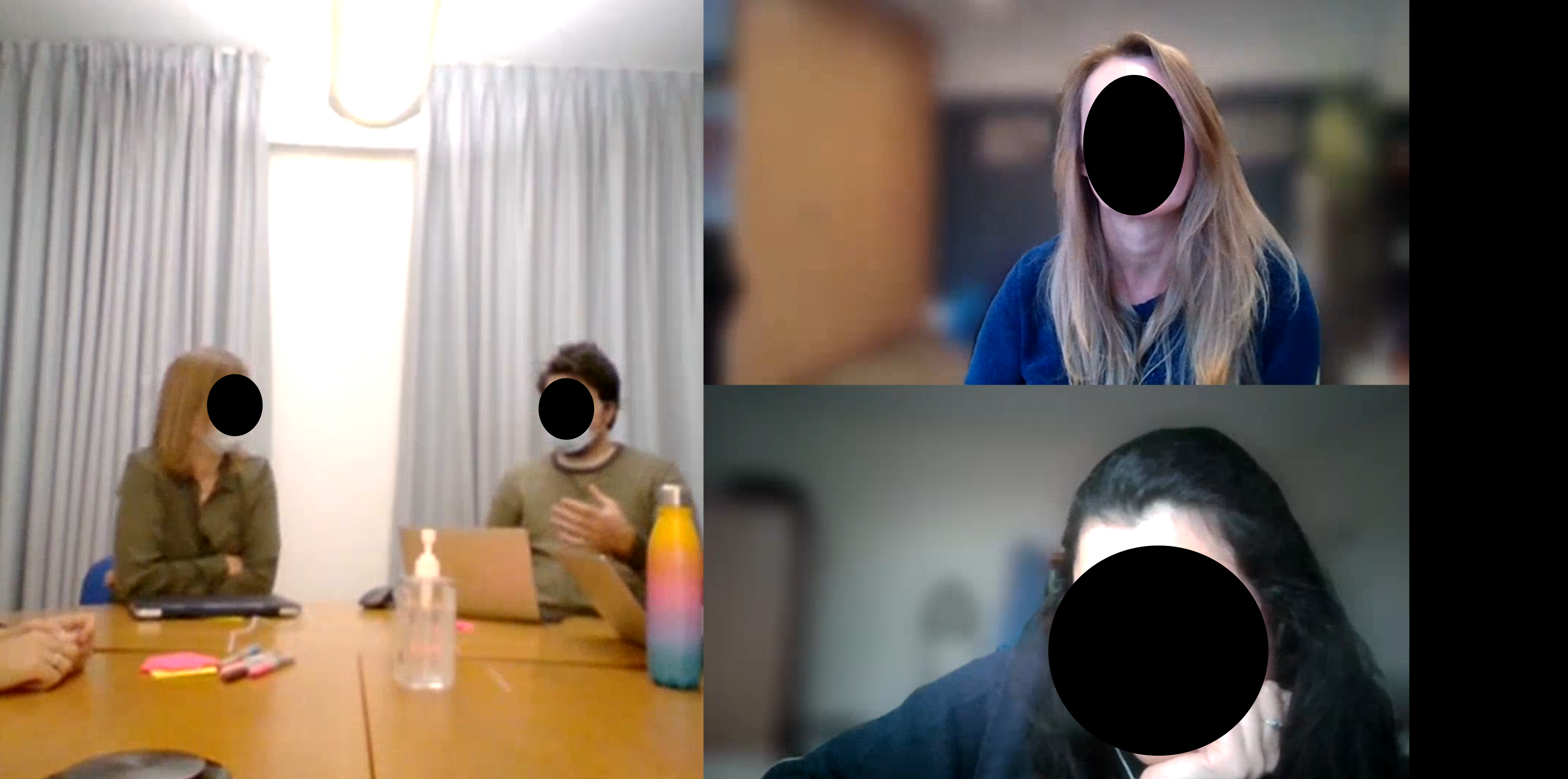}
    \caption{Two participants joined the focus group via MS Teams, and the other 3 (one of them is not depicted in the picture because she was out of the camera frame) joined in-person.}
    \label{fig:study3}
\end{figure}

\begin{figure}[h!]
    \centering
    \includegraphics[width = 0.7\textwidth]{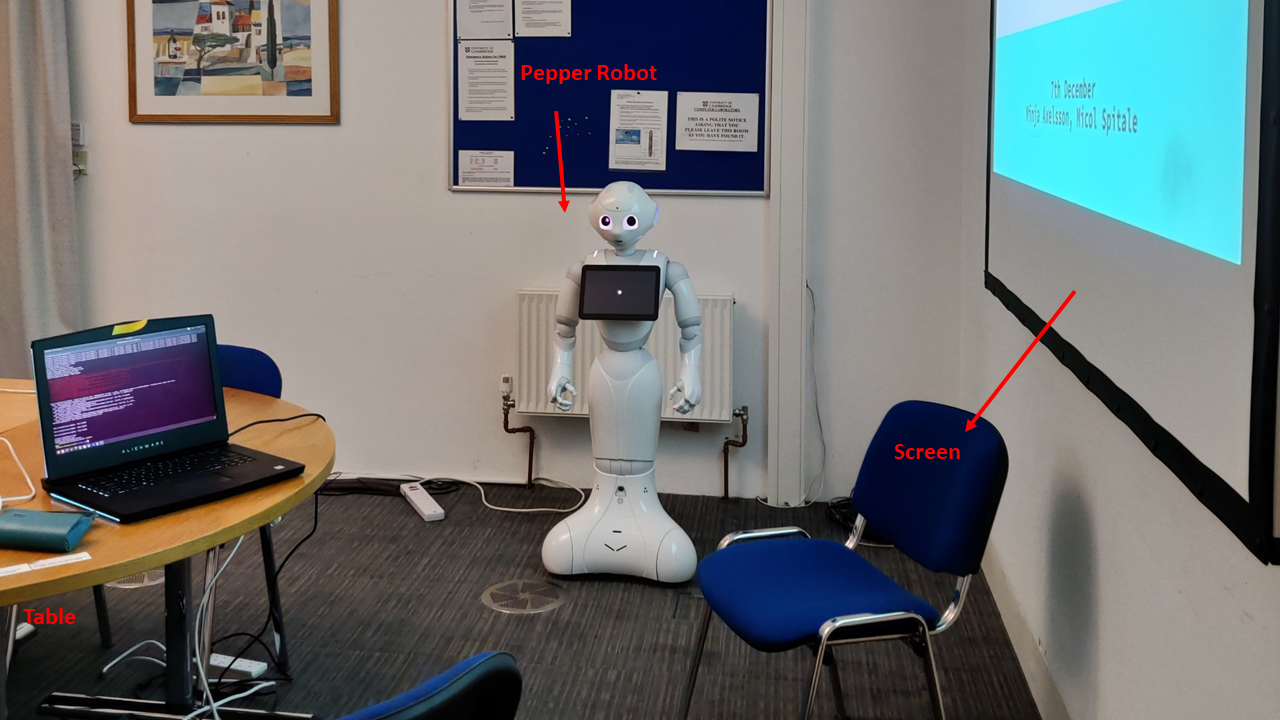}
    \caption{Setting of the focus group: a table around which participants were seated, a Pepper robot that performed a Positive Psychology interaction, and a screen that displayed the information required during the discussion.}
    \label{fig:setting}
\end{figure}

We conducted this study in a conference room of the University of Cambridge following an hybrid format. Three of the participants were present in-person while the other two joined the study remotely via MS Teams as shown in Figure \ref{fig:study3}. The in-person participants were seated around a table. The room was equipped with a projector that was used to display slides that contained information for the discussions, accessible to the remote participants in an online format. We placed the robot Pepper by the screen for a demonstration of the Positive Psychology exercises. Figure \ref{fig:setting} depicts the setting of the study.

\subsubsection{Protocol}

\begin{table}[h!]
    \centering
    \begin{tabular}{ll}
    \toprule
        \textbf{Items} &  \textbf{Duration}\\
        \toprule
         Demo of Positive Psychology with Pepper & 5 min\\
         Video of Mindfulness conducted with Pepper & 5 min\\
         Presentation of SFP by coach & 5 min\\
         Discussion after demo & 25 min\\
         Editing SFP script in groups & 45 min\\
         Ending discussion & 20 min\\
         \midrule
         Post-discussion questionnaires & 5 min\\
         \bottomrule
    \end{tabular}
    \caption{Structure of the focus group discussion}
    \label{tab:study3-structure}
\end{table}

The structure of the group discussion is detailed in Table \ref{tab:study3-structure}. First, we introduced ourselves and explained to the group the main goal of the session. Then, we asked one volunteer among the in-person participants to interact with a Pepper robot to demonstrate how the robot can deliver a Positive Psychology well-being session. All the participants watched the demo that lasted for about 5 minutes. 
The robot asked the participant what they were grateful for,
and what they had recently accomplished. 
A short discussion was held about the demo just experienced, prompted by questions we prepared (e.g., "Does the way the robot looks and talks feel right to conduct these exercises?") displayed on the screen for about 25 minutes.

After this, participants were also shown a video of Pepper conducting a Mindfulness session \cite{bodala2021teleoperated}. The coach present in-person introduced the concept of Solution-Focused Practice, as well as the main objectives of the practice, in order for all the participants to have initial knowledge on the topic. The coach answered the participants' questions related to the topic. The three demonstrations (Positive Psychology, Mindfulness, and SFP) were intended to give participants information about the range of well-being practices available, and to spark their imagination on what it would mean for a robotic coach to conduct them.

\begin{figure}[h!]
    \centering
    \includegraphics[width = 0.5\textwidth]{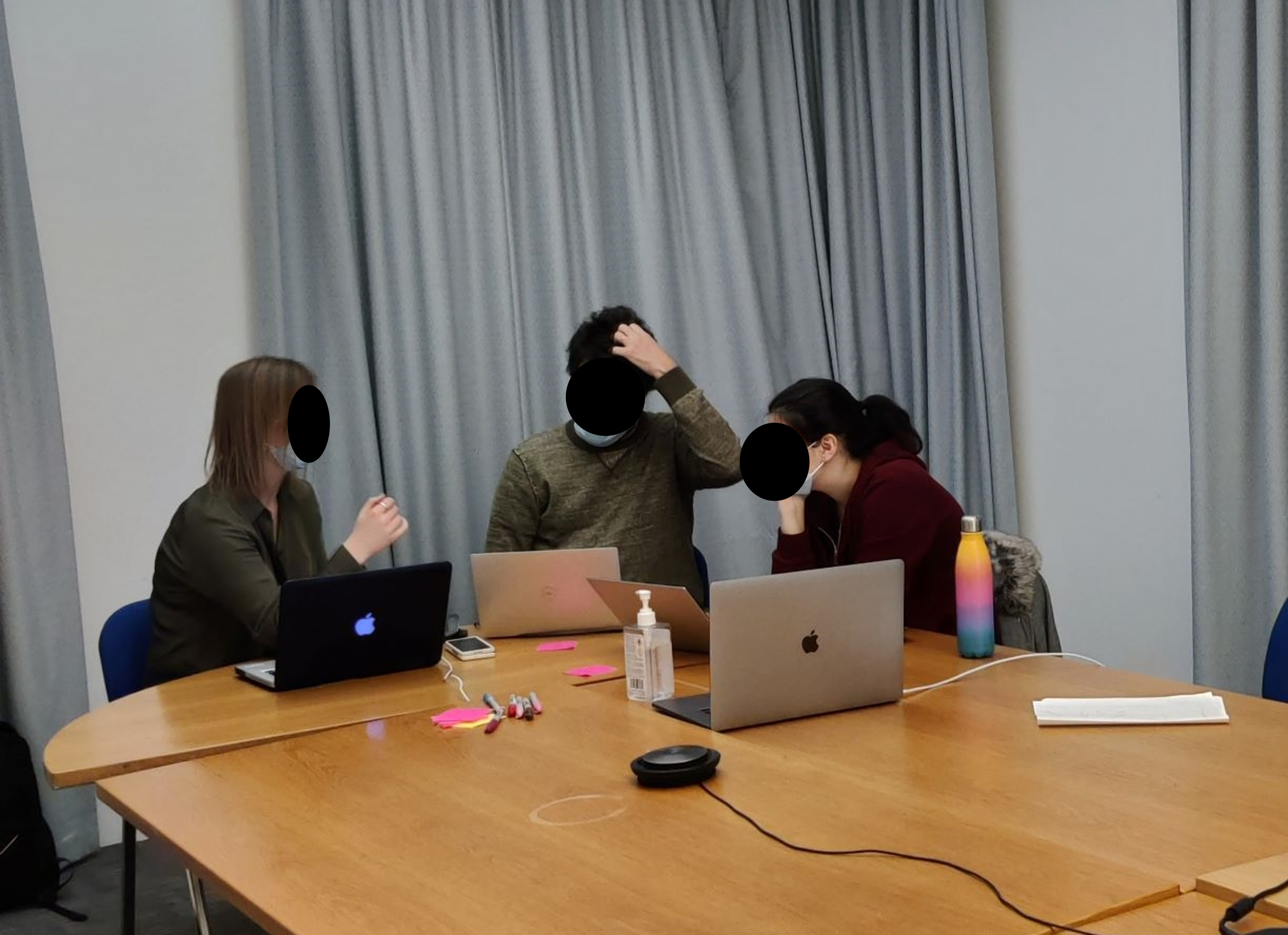}
    \caption{Participants are discussing the SFP script during their group activity.}
    \label{fig:groupdisc}
\end{figure}

We then split participants into two groups (one of 3, and the other of 2 people), and we assigned the well-being coaches respectively to each group. The members of each group were asked to collaborate with each other to discuss the use of robots in well-being practices, as shown in Figure \ref{fig:groupdisc}. Specifically, we provided them with a collaborative text document that contained a SFP dialogue script that had been transcribed based on sessions previously conducted by the SFP coach. Participants then edited this script based on how they would like a robotic coach to deliver SFP sessions (see the SFP script given to participants for editing in Appendix \ref{app:bsfp}). The group members were asked to edit the script and add comments to improve the sessions to be delivered by the hypothetical robot. We also supplied them with the online tool \textit{Miro} incorporating additional questions as well as portions of the Social Robot Co-Design Canvases \cite{axelsson2021social, Axelsson_2021}.

After 45 minutes, we began a group discussion on the topic of a robotic coach conducting SFP sessions, based on the edits the participants had made to the robot script. Again, we prompted them with a list of questions (e.g. "What would motivate you/your clients to start using a long-term robotic coach?") projected on the screen to facilitate the 20-minute discussion. Finally, we asked the participants to fill questionnaires to evaluate robot capabilities (described in detail in Section \ref{sec:survey3}).

\subsubsection{Surveys}
\label{sec:survey3}
We asked participants to fill out two surveys: 
a questionnaire with open-ended questions \hl{(e.g., "Refer to what you've put in the Miro board: What robot behaviours did you agree on with your pair?" or "Refer to what you've put in the Miro board: What 3 things would make you/your client most de-motivated to use a robotic mental well-being coach?")} to evaluate the motivation to start and keep using a robot well-being coach, and a demographic questionnaire.

\subsubsection{Data analysis}

We analyzed the data from the group discussions using TA following the method described in Study 1 (Section \ref{sec:study2}). 

\subsubsection{Main Findings}

\begin{figure}[h!]
    \centering
    \includegraphics[width=\textwidth]{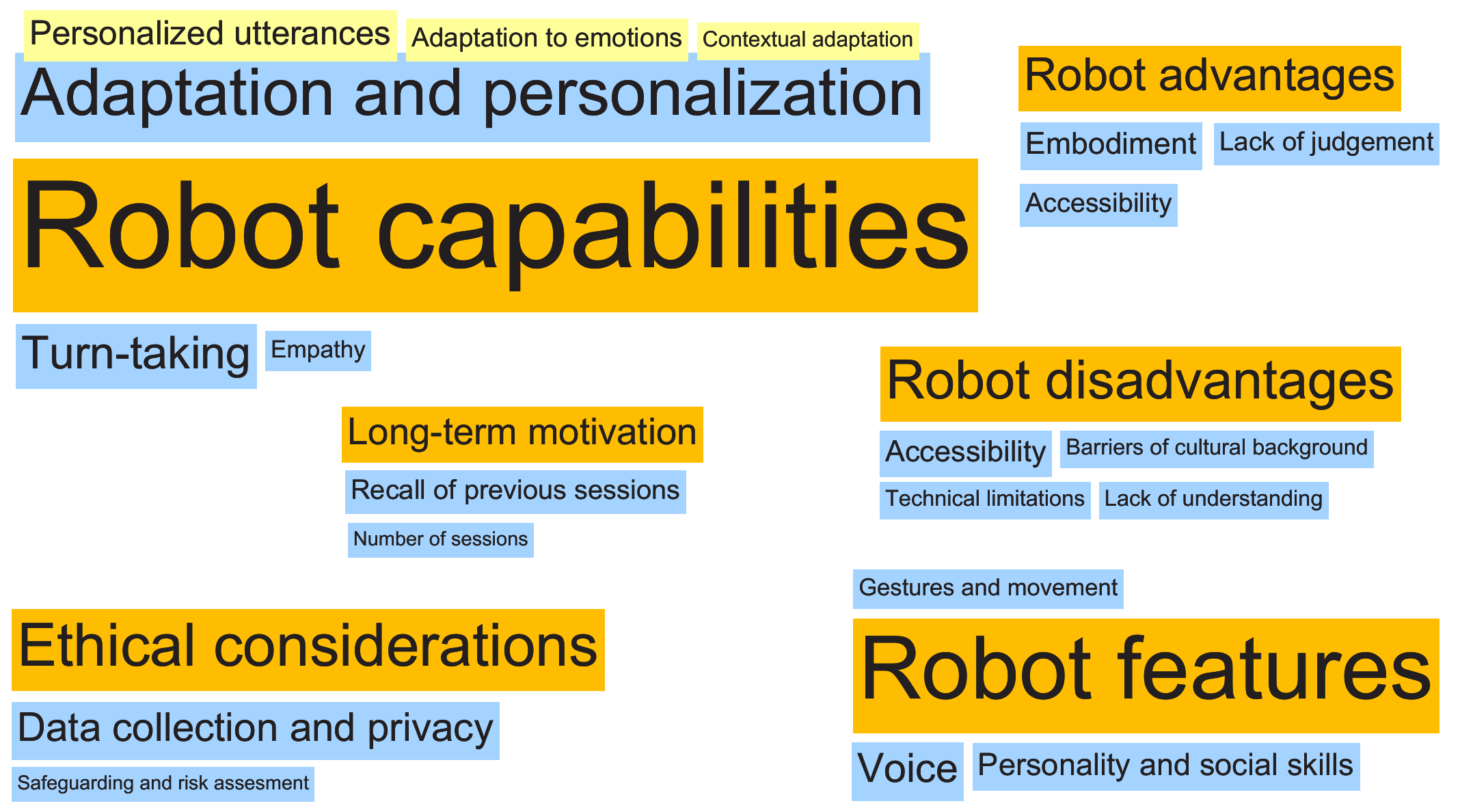}
    \caption{Study 3: Themes defined in the TA are presented in orange, while codes related to these themes are presented in blue, and sub-codes are presented in yellow (best viewed in colour).}
    \label{fig:study3TA}
\end{figure}

Figure \ref{fig:study3TA} depicts the themes identified in the thematic analysis. 
The following subsections report the themes and related findings, with quotes from the transcripts summarised where necessary (reported in Table \ref{table:study3_capabilities}, \ref{table:study3_adaptation}, \ref{table:study3_ethical}, \ref{table:study3_motivation}, \ref{table:study3_features}, \ref{table:study3_advantages} and \ref{table:study3_disadvantages}).

\begin{table*}[ht!]
\small
\centering
  \caption{Quotes from participants ($Pi, (i=1,2,3)$) and coaches ($Ci, (i=1,2)$) regarding \textit{robot capabilities} theme. }
  {
  \begin{tabular}{|C|L|}\Xhline{1.0 pt}
    \rowcolor{gray!40}
    \textit{\textbf{Robot capabilities}} & \textit{\textbf{Quotes from participants and coaches}}\\ \Xhline{1.0 pt}
    \rowcolor{gray!15}    

    Turn-taking
    & {
    P1: ``[...] you don't know if the pause is like, it's waiting for you to stop speaking so it can carry on, or if it's just you know, broken or something.’’
    
    P3: ``We don't understand whether our answers are being processed or not, and so there has to be some sort of cue that the robot is thinking or processing our information, and did not just malfunction and die.’’
    }\\
    \Xhline{0.5 pt}
    \rowcolor{gray!40}    
    
    Empathy
    & {C2: ``[...] I also feel at the moment Pepper is lacking compassion. Being there with the person and sharing in their suffering and empathising.’’
    
    C1: ``[Giving empathy at the person’s level], it's better [that the robot goes] just slightly above. Like if they're giving a little bit of hope, you just give a little bit more. You don't go wildly above, otherwise it feels really jarring. I mean when someone speaks quietly, we speak quietly. I think the robot could do that, would just [speak] slightly louder and I think that's what people do. Like you don't want your coach to sound depressed, but you also don't want them [to be jarring], so `I'm really jolly when you're down here', yeah?''
    }
   \\
    \Xhline{1.0 pt}
  \end{tabular}
  }
\label{table:study3_capabilities}  
\end{table*}

\paragraph{Robot capabilities}
Robot capabilities are seen in Table \ref{table:study3_capabilities}, as well as a separate table for the capabilities sub-theme \emph{Adaptation and Personalization} (Table \ref{table:study3_adaptation}).
Coaches and participants thought that the \textit{turn-taking} (i.e. improving signals of when the robot is listening and when it is about to speak) of the dialogue is crucial for a successful human-robot interaction. They reported that the feeling of being listened to and understood needs to be enhanced. They also believed that the gestures could be misleading and not clear enough indicators of turn-taking, suggesting that cues for this need to be designed further. Another important robot capability was conveying \emph{empathy}, as currently the robot was \emph{lacking compassion} \hl{as perceived by participants}. The SFP coach suggested empathizing \emph{at the person's level}, in order to avoid \emph{jarring} emotional expression. For example, if the person were feeling very low and speaking quietly, the robot should speak just slightly louder and give slightly more hope, rather than being highly expressive and joyful.

\begin{table*}[ht!]
\small
\centering
  \caption{Quotes from participants ($Pi, (i=1,2,3)$) and coaches ($Ci, (i=1,2)$) regarding \textit{robot capabilities: adaptation and personalization} theme.} 
  {
  \begin{tabular}{|C|L|}\Xhline{1.0 pt}
    \rowcolor{gray!40}
    \textit{\textbf{Adaptation and personalization}} & \textit{\textbf{Quotes from participants and coaches}}\\ \Xhline{1.0 pt}
    \rowcolor{gray!15}
    
    Personalized utterances
    & {C1: ``Maybe add some personalised comments in the middle about what the user said specifically?’’
    
    P3: ``Making it personalized, [...] picking on what people are specifically saying. [...] I don't want to know that my interaction with the robot was exactly the same, or [had] the same answers, as someone who had a completely different issue. So how can you make it personal?’’
    }
    \\
    \Xhline{0.5 pt}
    \rowcolor{gray!40}
    Adaptation to emotions
    &
    {C2: ``Just that the language should be adaptive and based on the responses of the participant, e.g. the response to a [mood score of] 2 should be very different from one to a 10.’’ 
    
    P2: `` I was thinking that we humans very rarely give a straight answer of 'I'm feeling happy'. Normally we'd balanced both [ends of the emotional spectrum] in the same sentence like 'oh, this thing happened'.’’
    }
    \\
    
        \Xhline{0.5 pt}
    \rowcolor{gray!15}
    
   Contextual adaptation
    & {C2: ``It would be great [for it to adapt to context] but I think this would be really tricky. I  would prefer that they respond and adapt to what the person is saying and their current mood/needs.’’

    P3: `Things like language  spoken, is it a special event, day, season, religious  context, location where the robot is based...’’ 
    }
   \\
    \Xhline{1.0 pt}
  \end{tabular}
  }
\label{table:study3_adaptation}  
\end{table*}

During the focus group, participants agreed on the fact that the most important features of a robot coach for well-being should be \textit{adaptation and personalization} to the user (see Table \ref{table:study3_adaptation}). 
We grouped adaptation and personalization as a single theme because participants used those terms as synonyms during the discussion. Coaches and participants believed that the robot should have \textit{personalized utterances} to each user to avoid the feeling of a ``mechanical'' interaction, \emph{adaptation to the user's emotions} for the coaching to be more successful, as well as \textit{contextual adaptation} to adapt to the context that the robot is being used in. Some of the participants pointed out that the robot should personalize the wording of its utterances to each user, even across sessions. However, the SFP coach highlighted that its personalization could be counter-productive in some cases, as SFP sessions have a set, scientifically evaluated and approved structure, that should be followed in order to attain the well-being benefits from the sessions. This indicates that the suggestions made by prospective users (who are laypeople uneducated in psychology) and professional coaches (educated in the underlying mechanisms of the well-being practices) may be in conflict. Careful evaluation is needed in determining when following each suggestion is appropriate, i.e. when to follow the coaches' suggestions to preserve coaching efficacy, and when to insert adaptation as suggested by the propsective users in order to create an experience of \emph{adaptation} and \emph{active listening} in the robotic coach. The desire for adaptation and personalization also conflicted with the participants' and coaches' concern about privacy and data collection, and they were apprehensive about storing data for long-term interactions. This is discussed further in Section \ref{sec:study3ethical}.

\paragraph{Robot features}

\begin{table*}[ht!]
\small
\centering
  \caption{Quotes from participants ($Pi, (i=1,2,3)$) and coaches ($Ci, (i=1,2)$) regarding \textit{robot features} theme.} 
  {
  \begin{tabular}{|C|L|}\Xhline{1.0 pt}
    \rowcolor{gray!40}
    \textit{\textbf{Robot features}} & \textit{\textbf{Quotes from participants and coaches}}\\ \Xhline{1.0 pt}

    \rowcolor{gray!15}
    Voice
    & {P1: ``I'll be quite surprised if Pepper's prosody will be able to make this long hypothetical bit [[in reference to Miracle Question in SFP]] sound natural.’’
    
    C1: ``Good point of it being hard to follow a long block of monotonous text from a robot, and [that you] can't ask it to repeat.’’
    }
   \\
    \Xhline{0.5 pt}
    \rowcolor{gray!40}
    Gestures and movement
    & {
    C2: ``[..] the hand gestures seemed fine, although I agree that a tilt of the head would be really natural as well.’’ 
    
    P3:  ``[it was] uncanny, I was like this is a robot, but the hands look too human. I don't know what to do with your movement. I did not understand.’’  
    }\\
    \Xhline{0.5 pt}
    \rowcolor{gray!15}
    Personality and social skills
    & {
    C2: ``I think pepper should use more encouraging, motivating language that reflects the person's strengths --- resilience, tenacity, desire to move things forward and change things about their lives, and willingness to engage in an exercise like this that is in their benefit.''

    P3: ``[...] making the robot have certain  characteristics means that the participant will feel empathy towards it, doesn't it? What if the robot describes being in an emotional state, that the  participant feels conflicting with their own state? [...] Perhaps friendly is a good personality?’’
    }
   \\

    \Xhline{1.0 pt}
  \end{tabular}
  }
\label{table:study3_features}  
\end{table*}

Participants discussed what features a robotic well-being coach should be endowed with (see Table \ref{table:study3_features}). The Mindfulness coach emphasized that the robotic coach should display \textit{empathic behavior} and compassion, and underline the user's \emph{strengths}.  Coaches and participants agreed that it would be important for users to feel that the robot was able to \emph{understand} them. 
Some participants noted that the synthesized voice of the robot lacked \emph{intonation and prosody}. They believed that the \textit{flat} and \textit{monotonous} voice could lead to an unnatural feeling during the interaction. They suggested that the robot should not say very long sentences in order to sound more natural.

Some participants experienced an uncanny perception during \textit{robot's gestures}, calling for more \emph{acknowledgement} gestures, i.e. nodding during the participant's speech. Participants believed that the robot's \textit{personality} should be neutral, but still warm and friendly. The robot should show positive behavior without being over enthusiastic during the coaching.

\paragraph{Robot advantages and disadvantages}

\begin{table*}[ht!]
\small
\centering
  \caption{Quotes from participants ($Pi, (i=1,2,3)$) and coaches ($Ci, (i=1,2)$) regarding \textit{robot's advantages} theme.}
   \label{table:study3_advantages}  
  {
  \begin{tabular}{|C|L|}\Xhline{1.0 pt}
    \rowcolor{gray!40}
    \textit{\textbf{Robot's advantages}} & \textit{\textbf{Quotes from participants and coaches}}\\ \Xhline{1.0 pt}
    \rowcolor{gray!15}

    Embodiment
    & {P3: ``What the robot does bring is an embodied [..]  it's the feeling that you're interacting with with a material form [..]  can bring it does bring a different reaction than if we were just speaking to a, uh, a TV or or something flat.’’
    
    C1: ``If it even vaguely looks human, they would be more likely to share [their] thinking, it's not like a device.’’
    }\\

    \Xhline{0.5 pt}
    \rowcolor{gray!40}
    Lack of judgement
    & {P1: ``Some clients might feel more at ease with technology than with other humans.''
    
    P2: ``I would in some cases fields more opens it, just tell pepper things that other humans 'cause they would have less judgement.’’
    
    C1:  ``Some of them might have preferred talking to a robot, knowing that it's not going to go anywhere like this. Yeah, there's no trust issue.’’
    
    P3: ``[...] caters to people uncomfortable with sharing their personal situation with other people.’’
    }\\
    
    \Xhline{0.5 pt}
    \rowcolor{gray!15}
    Accessibility
    & {C2: ``Not a person, but you know someone that you can talk to [...], has a friendly feel to it. I can imagine if I had one in my room I might well [...] get into the habit of chatting with it, but I don't know whether the benefits would be great enough versus a phone app, or a zoom call.’’
    
    P1: ``The robotic coach will also have more availability than a human.’’
    }\\
 
    \Xhline{1.0 pt}
  \end{tabular}
  }
\end{table*}

\begin{table*}[ht!]
\small
\centering
  \caption{Quotes from participants ($Pi, (i=1,2,3)$) and coaches ($Ci, (i=1,2)$) regarding \textit{robot's  disadvantages} theme. }
   \label{table:study3_disadvantages}  
  {
  \begin{tabular}{|C|L|}\Xhline{1.0 pt}
    \rowcolor{gray!40}
    \textit{\textbf{Robot's disadvantages}} & \textit{\textbf{Quotes from participants and coaches}}\\ \Xhline{1.0 pt}
    \rowcolor{gray!15}

    Technical limitations
    & {C2: ``[..] prompting people to reflect on what they're grateful for, is really beneficial. I'm not sure for me what Pepper is adding to that process, versus an app on your phone that might just pop up at a certain time and say, tell me what you're grateful for. [..] What does Pepper bring that is beyond just the auditory stimulus to reflect?’’
    
    P1: ``I think that there is still quite a lot of programming and technological [work to be done] [...]. It's increasing the type of reactions, understanding the words people are saying, not requiring people to repeat themselves, giving some sort of continuing body movement. I agreed, the hands were very creepy when I did the exercise.’’
    
    P3: ``Maybe it doesn't make any difference that it's a robot, or we could do the same by any other virtual means.’’
    }
    \\
    \Xhline{0.5 pt}
    \rowcolor{gray!40}
    Lack of understanding
    & {C1: ``You're having to repeat, and repeat loudly, when you're doing an exercise [in order for the robot to hear you].’’
    
    P2: ``It doesn't matter what we say to the robot, but perhaps it matters what I say to myself.’’
    }\\
    
     \Xhline{0.5 pt}
    \rowcolor{gray!15}
    Barriers of cultural background
    & {P3: ``[...] it touches upon all of these very tricky feelings and emotions. That in a lot of cultures, they're not to be openly expressed. Even to a robot, [...] you don't share it. And so the robot works in specific contexts, it does. But then in other sociocultural contexts, the robot will not be able to. People won't give the robot the information that they're supposed to give. It's not culturally acceptable to do that.''

    C2: ``[...] we just need to make make it clear that it may not be adaptable to other cultures and people.''
    }\\
    
    \Xhline{0.5 pt}
    \rowcolor{gray!40}
    Accessibility
    & {P2: ``[..] money issues.’’
    
    C2: ``The barrier to using it would be the cost and availability and access to Pepper, which would mean that the benefits would have to be really quite big [...]’’
    }
   \\
    
    \Xhline{1.0 pt}
  \end{tabular}
  }
\label{table:quotes4}  
\end{table*}

Coaches and participants thought the robot might have certain advantages (Table \ref{table:study3_advantages}) and disadvantages (Table \ref{table:study3_disadvantages}), particularly discussing these in comparison to a \emph{mobile app} or a \emph{human coach}. Some participants saw the robot's \emph{embodiment} as an advantage in comparison to a mobile app, while the Mindfulness coach and some participants saw that the \emph{barriers of the technology} were limiting what the robot could accomplish, and thus would not see much difference in using the robot in comparison to a mobile app. The robot's \emph{lack of understanding} was also seen as a limitation, in particular the SFP coach remarked that people would have to sometimes repeat themselves loudly when talking to the robot. However, one participant remarked that ``It doesn't matter what we say to the robot, but perhaps it matters what I say to myself'', indicating that the robot could be useful even with limited understanding. 

The robot's \emph{lack of judgement} was seen as an advantage, as it could help users \emph{trust} the robot more. Participants noted that people who were uncomfortable with sharing personal situations with other people might find the robot useful. On the other hand, barriers related to \emph{cultural background} could reduce this advantage. P3 noted that in some cultures, emotions are not openly expressed, and this could apply also to expressing them to a robot. As noted, the deployment of such robots should be carefully considered in light of the cultural context of its environment.

\emph{Accessibility} of the robot was seen as both an advantage and disadvantage. Some of the participants believed that cost and availability are contributors to the unpopularity of robots currently, and that they were unsure whether a robot's benefits would outweigh such barriers. On the other hand, some participants noted that having the robot in their room would encourage them to use it, and that a robotic coach could have more availability than a human coach.

\paragraph{Motivation over time}

The group was also prompted to discuss what would make them motivated to use a robotic well-being coach over longer periods of time (see Table \ref{table:study3_motivation}). Interestingly, the SFP coach clarified that the duration of SFP coaching actually depends on the coachees, and how many sessions they needed and benefited from. The coach mentioned that the coaching has succeeded when it is no longer needed. For example, some people could need only a single session to feel better, and others could need 10 sessions before considering the well-being coaching successful. However, other well-being practices such as Mindfulness practices are different in that they can be practiced every day, without the intervention targeting a particular problem, and can continue indefinitely.
Participants discussed that in the case users wanted multiple sessions, i.e., a longitudinal interaction with the robotic well-being coach, an important robot capability is the robot's memory of the past sessions. The robot should recall how the users felt during past sessions, because it could be beneficial and rewarding for the user to check their progress. This would also help facilitate the perception that the robot was \emph{acknowledging} and \emph{actively listening} to the participants. Participants also stated that the robot should avoid \textit{repetitive behavior} in long-term interaction in order to avoid the perception of the robot being mechanical in nature, which might create frustrations in the user.

\paragraph{Ethical considerations}

\begin{table*}[ht!]
\small
\centering
  \caption{Quotes from participants ($Pi, (i=1,2,3)$) and coaches ($Ci, (i=1,2)$) regarding \textit{ethical considerations} theme.}
  {
  \begin{tabular}{|C|L|}\Xhline{1.0 pt}
    \rowcolor{gray!40}
    \textit{\textbf{Ethical considerations}} & \textit{\textbf{Quotes from participants and coaches}}\\ \Xhline{1.0 pt}
    \rowcolor{gray!15}
    Data collection and privacy
    & {C2: ``We need to be very transparent, very upfront, and the conversation about you know what, if anything is being recorded and transmitted.’’
    
    P1: ``I didn't want to share something extremely intimate [in a group setting with the robot]. Yeah, but I was trying to say something that makes sense as an answer, but it's not at the same time exposing myself too much.’’
    }
   \\
    \Xhline{0.5 pt}
    \rowcolor{gray!40}
    Safe-guarding and risk assessment
    & {
    C2: ``There may be at risk of suicidal ideation or self harm or and I just feel that [...] we need to risk assess first before we get into the issues that that person wants to talk about it.’’
    
    C2: ``If Pepper comes across someone who is in crisis, either suicidal, self-harming, or in danger. What are the safeguarding processes? Is there a way Pepper can check if people are at risk before the conversation starts?'' 
    }\\
    
    \Xhline{1.0 pt}
  \end{tabular}
  }
\label{table:study3_ethical}  
\end{table*}

\label{sec:study3ethical}

Participants discussed several ethical considerations related to robotic well-being coaches, as presented in Table \ref{table:study3_ethical}. 
The participants' main concern when interacting with a robotic well-being coach was \textit{data collection and privacy}.
During the discussion, the Mindfulness coach highlighted that \emph{transparency} is a key factor for a successful coaching session: the robot should state upfront if it is collecting or recording any data, in compliance with GDPR (the General Data Protection Regulation of the European Union \cite{gdpr2022}). However, transparency could also affect the efficacy of coaching. The volunteer participants during the Positive Psychology session noted that they were considering what they should share with the robot, as they were conscious that other people could hear them. One of the discussion groups noted that privacy could especially be a concern in countries with totalitarian regimes, where expression of negative emotions or concerns is not socially accepted, and may even be surveilled. This indicates the importance for cultural consideration when deploying such robots, and the responsibility of roboticists to ensure that sensitive data is appropriately protected.

Another ethical consideration highlighted by the well-being coaches was the importance of \textit{assessing risk and safeguarding} during well-being coaching. Coaches expressed that the robot would not be appropriate to be used with clinical patients experiencing severe distress (e.g. suicidal ideation). The discussion converged in the viewpoint that users of such a robot should be pre-screened by human practitioners for any warning signs, as a robot may not be able to assess these types of situations.

\begin{table*}[ht!]
\small
\centering
  \caption{Quotes from participants ($Pi, (i=1,2,3)$) and coaches ($Ci, (i=1,2)$) regarding \textit{motivation over time} theme.}
   \label{table:study3_motivation}  
  {
  \begin{tabular}{|C|L|}\Xhline{1.0 pt}
    \rowcolor{gray!40}
    \textit{\textbf{Motivation over time}} & \textit{\textbf{Quotes from participants and coaches}}\\ \Xhline{1.0 pt}
    \rowcolor{gray!15}
    Number of sessions
    & {C1: ``Interventions should be as brief as they need to be or can be, so I wouldn't want it to be used indefinitely. [...] It depends on the person. A person could need a single session or 100 sessions.’’
    }
   \\
    \Xhline{0.5 pt}
    \rowcolor{gray!40}
    Recall of previous sessions
    & {C2: ``There's no reminders of how well I'm doing or how positive it is that I'm engaging in this thing --- in this process that could be to my benefit [...]. [Thinking] 'I am doing quite a good job', 'I'll keep doing it', so [adding more] strength language in there.’’
    
    P1: ``Remembering what happened previously and saying, 'oh, you are feeling better than you were last week'?''
    }\\
    \Xhline{1.0 pt}
  \end{tabular}
  }
\end{table*}

\section{Qualitative Analysis: convergence and divergence of the studies}
\label{sec:meta-analysis}

The three studies presented in this paper converge and diverge in different themes defined with the TA. \textcolor{black}{As all studies were part of an iterative design process, each study built on the conclusions of the previous ones. Due to the flexible and open nature of the study procedure, differences and convergences emerge between the feedback gathered from the stakeholders.}  
In the next sections, we will examine the \emph{convergence} (i.e. similarities) and \emph{divergence} (i.e. differences between) study results, in order to highlight both complementary findings, as well as points of departure and the reasons for these departures \cite{timulak2014qualitative}. All themes and their occurrence across the studies are presented in Appendix \ref{app:codebook}.

\begin{table}[ht!]
  \caption{Comparison of 3 studies}
  \label{table:study-comparison}
  \begin{tabular}{A{1,2cm} A{1,8cm} A{1,2cm} A{4,3cm} A{3,0cm}}
    \toprule
     & {\textbf{WBP}} & {\textbf{Coach}} & {\textbf{Duration}} & {\textbf{Data collection}} \\
    \toprule
    {\textbf{Study 1}} & {BSFP; Mindfulness} & {Human, first person} & {Long-term (four 20 min sessions over 4 weeks)} & {Individual interviews, group discussions}   \\
    \midrule
    {\textbf{Study 2}} & {Positive psychology} & {Robot, first person} & {Short-term (one $30 \pm 11$ min session)} & {Individual interviews} \\
     \midrule
    {\textbf{Study 3}} & {Positive psychology; Mindfulness; BSFP} & {Robot, first/third person} & {\textbf{PP:} Short-term (one 10 min observed session), in some cases took part in Study 2 as well; \textbf{Mindfulness:} Video demonstration; \textbf{BSFP:} Introduced by coach} & {Group discussion}  \\
    
  \bottomrule
\end{tabular}
\label{table:study-comparison}
\end{table}

\subsection{Convergence}

\label{sec:convergence}

All three qualitative studies detailed similar capabilities and features of a robotic well-being coach, as well as what advantages and disadvantages it may have, especially in comparison to a human coach or a mobile app delivering the same exercises. This section details which of the sub-themes within these themes were expressed similarly across studies.

\subsubsection{Robot capabilities}
\label{sec:convergence_capabilities}
\mbox{}\\

\textbf{Acknowledgement and active listening} ---
Study 2 and Study 3 emphasized the robotic capabilities of \emph{acknowledgement} and \emph{active listening}, and how these capabilities could be improved in order to enhance the user's experience and \emph{feeling of being listened to}. Participants noted that these capabilities could be expressed through \emph{verbal utterances} by using \emph{wording variety}, \emph{keywords} from the participants' speech, \emph{paraphrasing} and \emph{summarising}, as well as \emph{phatic expressions}. \emph{Body language} could also be used to convey \emph{backchanneling}, with the robot nodding, leaning in, or tilting its head.

\textbf{Expressing empathy} ---
\label{sec:empathy_expression}
All three studies mentioned the importance of the robot \emph{expressing empathy}, with Study 3 specifically going into detail on what level of empathic expression is appropriate (see Section \ref{sec:divergence_emotional_adaptation}). Empathy was seen to be related to the \emph{feeling of being listened to}, and could on a more sophisticated level include \emph{emotional adaptation}.

\textbf{Adaptation and personalization} ---
\emph{Adaptation and personalization} were considered important in all three studies, with Study 3 especially going into detail on \emph{personalized utterances}, \emph{adaptation to users' emotions}, and \emph{contextual adaptation} (see Section \ref{sec:study3}). Participants felt that adaptation and personalization (terms which most of them used interchangeably) could help them \emph{engage} in the practice. However, the appropriate level of this adaptation was not agreed upon by all participants (see Section \ref{sec:divergence_emotional_adaptation} for discussion).

\subsubsection{Robot features}
\label{sec:convergence_features}
\mbox{}\\

\textbf{Form} ---
Robot features were discussed in a broad and hypothetical sense (without reference to any specific robotic implementation) in Study 1, and more specifically in Study 2 and Study 3, where there was a specific robotic application to evaluate and improve. While the preferences for the robot's \emph{form} --- i.e. humanoid and abstract robots --- diverged in Study 1 (see Section \ref{sec:divergence_form} for discussion), the humanoid robot Pepper received mainly positive feedback during Study 2 (with some participants still noting they would prefer a more abstract robot). 

\textbf{Voice and movement} ---
The robot's \emph{voice} and \emph{movements} were mainly discussed during post-interaction interviews in Study 2. Pepper's default voice, which was used during these studies, received positive feedback for its slow pace and friendly tone, and the movements were generally received positively, but needed improvement with regards to \emph{expressing acknowledgement} and \emph{less repetition}. Additionally, participants in Study 3 noted that Pepper's default voice could be slower and use more variation in prosody, in order to better engage the user and be more understandable.

\textbf{Behaviour} ---
The robot's \emph{behaviour} was mainly examined via post-interaction interviews in Study 2, and via group discussion in Study 3. In both studies, participants were focused on improving the capabilities of \emph{responsiveness} and \emph{adaptation} in the robot's behaviour (see Section \ref{sec:convergence_capabilities}). Outside of these capabilities, participants in Study 1 discussed the robot's behaviour from the perspective of being \emph{too forceful or patronizing} versus \emph{not being direct enough}. In Study 2, the robot's behaviour was widely regarded as \emph{friendly} and \emph{supportive}, with only a few participants mentioning that the robot could at times come off as patronizing. Some participants in Study 2 found it positive that the robot did not express too much emotion. In study 3, participants also emphasized that the robot should be friendly and compassionate, as well as not too emotionally expressive.

\subsubsection{Robot advantages}
\label{sec:convergence_advantages}
\mbox{}\\

\textbf{Lack of judgement} ---
The robot was also seen to have a \emph{lack of judgement} in comparison to a human coach, across all three studies. In Study 1, two coaches noted that the robot would have this as an advantage over people, with the Life Coach noting that their life experiences could bias them to attempt to solve the coachee's problems in a non-productive way. Participants in Study 2 noted that the robot Pepper not having facial expressions was actually a strength, as it could take in what the participant was saying without reacting and being judgemental. In Study 2 and Study 3, participants brought up that users who were not comfortable sharing their feelings or thoughts with other humans could make use of robotic coaches. \hl{Future work should explore how in particular introverted or socially isolated people could benefit from a robotic coach.}

\textbf{Presence and embodiment} ---
The robot advantage of \emph{embodiment}, i.e., \emph{having a presence}, was seen in both Study 1 and Study 3. Participants noted that having something physically present, instead of using, e.g., a mobile app, could help participants feel \emph{more engaged} in the well-being practice. In Study 1, participants noted that the presence of a robot would also act as a \emph{visual reminder} for the practice. In Study 2, participants noted that one of the major advantages of the robot in comparison to a mobile app was \emph{talking out loud}. Since the participants regarded the robot as a conversation partner, they noticed that even saying their thoughts out loud was helpful, even if the robot did not always understand what they said. This same sentiment was expressed by a participant in Study 3.

\subsubsection{Robot disadvantages}
\label{sec:convergence_disadvantages}
\mbox{}\\

\textbf{Lack of humanness} ---
\emph{Lack of humanness}, i.e. a robot's lack of human-like communication, was a disadvantage mentioned in Study 1 and 2. This is the other side of the picture with the robot's advantage of \emph{lack of judgement} --- while a robot can provide a judgement-free well-being platform, sometimes those reactions can be desired and even deemed necessary by participants. Specifically, participants wanted to be able to dynamically correct \emph{incorrect responses} of the robot during conversation, i.e. resolve misunderstandings. Additionally, during Study 2, participants wanted to be able to \emph{ask for clarification}, as well as \emph{receive suggestions} based on their answers to the robot's questions. While these capabilities could be added to a robot, the level of intuitiveness that people have (e.g., the coach being able to recognize a look of confusion on a person's face, and providing clarification due to this) may not be achievable with a robot. This is related to Study 3, where participants noted that \emph{technical limitations} could demotivate them from using a robotic mental well-being coach.

\subsubsection{Ethical considerations}
\label{sec:convergence_ethical}
\mbox{}\\

\textbf{Privacy and data collection} ---
\emph{Privacy} was the main perceived ethical concern a with robotic mental well-being coach in Study 1, 2 and 3. Especially in Study 2, a participant with robotics experience noted that they were conscious of what experiences they were sharing with the robot since they knew the data would be reviewed by human researchers later. The same participant repeated the same concern in Study 3. During Study 3, the Mindfulness Coach emphasized that the robot should not collect or store any data --- however they also said that the robot should be able to adapt its behaviour to each participant according to their wishes. Due to this conflict, the disadvantage of privacy concerns is particularly tricky, as participants may not understand that full preservation of privacy and capabilities wished for from the robot may be incompatible. On the contrary, the SFP Coach involved in Study 1 noted that a robot may actually provide more privacy than a human, as in some situations a human coach may be required to disclose important information to other people, or may disclose something due to human error (as also noted by the Life Coach in Study 1). To mitigate this concern, informing participants of what data is stored and how it is used could be applied. This is discussed further in section \ref{sec:design}.

\textbf{Safeguarding and risk assessment} --- Safeguarding and risk assessment came up as an ethical consideration in Study 1 and Study 3. In Study 1, participants noted that the robot should have a protocol for if a user were in distress. In Study 3, the Mindfulness coach in particular noted that users should ideally be pre-assessed before interacting with the robot, and directed to other resources if they were in crisis. The coach and some participants remarked that the robotic coach would only be appropriate for people who did not have significant mental health challenges, and were rather seeking to improve their well-being.

\subsection{Divergence}
\label{sec:divergence}

Divergences in themes between the studies may be due to the different framing of all of the studies (i.e. different topics of discussion, different data collection methods, different well-being practices being discussed). Divergences between participants' opinions also emerged within the studies, possibly as a result of different perspectives (participant versus coach) or different backgrounds (SFP coach versus Mindfulness coach, or participant in technology field versus anthropology field). Additionally, the three studies had different approaches \textcolor{black}{since they were conducted in consecutive phases of the iterative design process}: Study 1 was a hypothetical, broader discussion of the potential of robotic mental well-being coaches; Study 2 examined a particular implementation of Pepper as a Positive Psychology coach; Study 3 gave an overview of three robotic well-being coaches and asked participants to directly edit a script for a SFP robotic coach, and to discuss their opinions (see Table \ref{table:study-comparison}). This section first reviews the divergences in opinion, and then the divergences in the emphasis of different themes due to the different framing of studies.

\hspace{5pt}
\\
\textbf{Divergences in opinion}

\subsubsection{Robot form}
\label{sec:divergence_form}

Robot form was discussed in detail in Study 1, while in Study 2 and 3 the robot's form was assumed to be Pepper. In Study 1, participants disagreed with each other about the robot's form, with some preferring a more abstract robot and others a more humanoid robot (see Section \ref{sec:study1}). However, all agreed that the robot's form should follow its function --- i.e. that they would expect a more humanoid robot to have more human-resembling conversational capabilities. 

While participants had differing opinions on preferred robot form in Study 1, Pepper's form was largely accepted during Study 2 and Study 3, where it was not explicitly a topic of discussion. A few participants called for a more abstract robot during Study 2, and some had specific criticisms of Pepper's form (the hands were creepy, or the light in Pepper's eye was a distraction). However, participants did note during Study 2 that they desired more human-like conversational qualities from the robot (i.e. responsiveness and adaptation) due to its human-like form. Similarly, while the form of Pepper was not explicitly discussed during Study 3, participants focused mainly on designing responsiveness and adaptation features for a robotic coach, after being presented the robot Pepper as an interaction platform.

\subsubsection{Adaptation and privacy}
\label{sec:divergence_adaptation_privacy}

During Study 3, the conflict between people wanting adaptation and privacy simultaneously was discussed in detail. This concern became apparent when participants were editing the SFP script of the robot, helping them understand that wherever they wanted the robot to adapt or personalize specifically to data they had supplied during the current session or the previous session, the robot would have to record such data. This concern was not discussed in Study 1, as the study was more about a hypothetical robot. In Study 2, participants also called for more adaptation, and while they were concerned about privacy, they did not state as strongly that the robot should not collect any data. Rather, they emphasized that the participant should be well informed about what data is collected (which had taken place before the study).

\subsubsection{Emotional adaptation}
\label{sec:divergence_emotional_adaptation}

In Study 1 and Study 3, the SFP coach held the view that it would be better for the robot to not adapt to a person's emotional experiences too much, due to the risk of being jarring if the robot got the reactions wrong. The coach held that the robot could be helpful even with minimal emotional adaptation, while retaining a \emph{friendly and neutral personality}. However, during Study 3, the Mindfulness coach as well as some participants said that the robot in its current state was not compassionate enough, and more \emph{adaptation to emotions}, as well as \emph{expressions of empathy}, were necessary for the robot to be useful. These participants had \emph{expectations} toward a robotic coach that can be difficult to achieve due to \emph{technical limitations}. Participants expected \emph{personalized utterances}, while the SFP coach held that these are not necessary for certain types of well-being coaching to work. During Study 2, several participants noted that the robot's emotional expressions were good when they were correct, however they felt they were jarring when the robot assessed their emotions incorrectly. These data indicate that it is difficult to determine the correct level of emotional adaptation, as the robot will inevitably make some mistakes. This is discussed further in Section \ref{sec:design}.

\subsubsection{Accessibility}
\label{sec:divergence_accessibility}

Accessibility of a robotic mental well-being coach was seen as both an advantage and disadvantage, particularly in Study 3. In Study 1 and Study 3, participants pointed out that a robotic well-being coach could have more availability than a human coach, with the SFP and Life Coaches in Study 1 remarking that the robot could also be more \emph{neutral} due to lack of daily routines or past life experiences. However, during Study 3, participants mentioned that the cost of a robotic coach such as Pepper could be quite high, and the benefits would have to be high enough to justify that cost. In Study 1, participants remarked that they could see themselves using such a robot if it were easily accessible in a space such as their home or their workplace. However, these options may not be available to all people. Accessibility (as well as the acceptability) of such a robotic coach will depend on the socio-economical and cultural contexts, and should be examined on a case-by-case basis.

\hspace{5pt}
\\
\textbf{Divergences in theme emphasis}

\subsubsection{Well-being practices and robot roles}
\label{sec:divergence_wellbeing_practices}

Study 1 focused explicitly on what well-being practices a robotic coach could perform, and what it should and shouldn't do in relation to these. Study 2 focused on giving feedback to a robotic coach conducting a \emph{Positive Psychology} session specifically, and Study 3 focused mainly on editing an SFP script. Due to this, \emph{robot roles} were also discussed more in Study 1, since they relate directly to how a robot should conduct itself --- i.e., what role it should assume --- when delivering well-being practices. During Study 1, well-being practices such as yoga and life-coaching practices were discussed, which weren't discussed during the later studies, which focused on Positive Psychology and SFP specifically. The robot was also seen to be able to for example take the role of a practice buddy during yoga (i.e., doing the practice with the participants), while this robot role did not come up during discussions regarding Positive Psychology and SFP. 

\subsubsection{Broader advantages}
\label{sec:divergence_broader_advantages}

Study 1 was a broader, hypothetical examination of potential robotic coaches for mental well-being. As such, some of the potential advantages of a robotic coach that were expressed in Study 1, were not discussed or seen to be expressed in Study 2 or Study 3, where a robotic application was examined in detail. Advantages that did not receive much discussion in Study 2 or 3 were \emph{neutrality and anonymity} and \emph{reliability and consistency}. \emph{Memory} and \emph{data analysis} were not mentioned in Study 2, which may be due to the one-off nature of the interaction, and these are features that would be visible only in longitudinal interactions. In Study 3, participants noted that they would like to see the robot \emph{recalling previous sessions} and \emph{adapting to the users' emotions}, both previous and current (see Tables \ref{table:study3_adaptation} and \ref{table:study3_motivation} ).

\subsubsection{Phenomena related to HRI}
\label{sec:divergence_phenomena}

\emph{Novelty and habituation effects} as well as \emph{suspension of disbelief} were phenomena related to HRI observed in the themes of Study 2 (see Section \ref{sec:study2_phenomena}). As Study 1 did not involve HRI, and Study 3 did not explicitly analyze the HRI itself, these phenomena were not observed in those studies. During the Positive Psychology exercises conducted by the robot Pepper, participants noted that their experience with the robot changed throughout the interaction session, and that they had to \emph{get used to}, i.e., habituated to, the robot. Participants noted initial feelings of strangeness, and feeling conscious about the experience, which related to both the \emph{novelty} effect as well as the lack of \emph{suspension of disbelief}.

\section{Design and ethical recommendations}
\label{sec:design}

As the result of our \textcolor{black}{qualitative analysis} of the three studies described in Sections \ref{sec:study1}, \ref{sec:study2} and \ref{sec:study3}, we present design guidelines for robotic well-being coaches, selected on the basis of the convergence and divergence of the findings in these studies. The guidelines address the robot's \emph{form}, its \emph{voice}, as well as the role of \emph{user and robot personality}, how the robot should \emph{acknowledge} and \emph{actively listen} to the user, the role of \emph{turn-taking}, the level of \emph{verbal adaptation}, other \emph{adaptation and personalization}. Additionally, we introduce ethical considerations that we think are particularly important when designing robotic coaches for well-being. The guidelines and ethical considerations, collected in Table \ref{table:design-guidelines-rationale} and \ref{table:design-guidelines-rationale2},  were selected to operationalize the results that converged across the three studies, while addressing divergences of results and attempting to resolve these conflicts. \hl{These recommendations are given within the context of current robot capabilities and research into robotic well-being coaches. Some recommendations may also be applicable in robots deployed in other contexts, e.g., universities, workplaces, or homes. However, these contexts are out-of-scope for this paper, and researchers should evaluate how each recommendation applies to their robotic coach within their context.}

\begin{table}
  \caption{Design and ethical recommendations rationale}
  \label{table:design-guidelines-rationale}
  \begin{tabular}{A{2,5cm} A{4,2cm} A{5,9cm}}
    \toprule
      {\textbf{Recommendations (HOW)}} & {\textbf{Issue (WHAT)}} & {\textbf{Rationale for recommendations (WHY)}}  \\
        \midrule
    \multicolumn{3}{l}{\textbf{Category: Robot Form}} \\
    \midrule
    {\textbf{(R1)} Robot form should follow its function} & {The robot's appearance (i.e. humanoid vs. abstract) should match its level of sophistication in its function (e.g. conversational skills).} & {Participants mentioned this being a requirement in Study 1 (Section \ref{sec:divergence_form}). Form and function attribution has been previously examined in HRI design \cite{haring2018ffab}. As different robot forms are matched to different levels of well-being interactions (e.g. humanoid for conversational, pet-like for care-taking), this should be kept in mind.
    }    \\
    \midrule
    {\textbf{(R2)} Robot voice should emphasize variable prosody and slow pace} & {Prosody, as well as slow pace, will be particularly important for explanation-heavy (e.g. the Miracle Question in SFP) or instruction-based (e.g. Mindfulness and Meditation) well-being practices.
    } & {Participants regarded the robot's voice mainly positively (Section \ref{sec:convergence_features}), but during Study 3, participants noted that Pepper's default voice should be slower and use more prosody, in order to engage the user and be understand (Section \ref{sec:study3}). A monotonous voice is more difficult to focus on. Additionally, a slow pace is often used in Meditation to induce calmness \cite{knowlton2006influence}.
    } \\
    \midrule
    \multicolumn{3}{l}{\textbf{Category: Robot Behaviour (Adaptation and personalization)}} \\
      \midrule
    {\textbf{(R3)} Acknowledgement and active listening should be implemented through backchanneling and appropriate turn-taking
} & {Backchanneling can be done through body language such as nodding, and phatic expressions. The robot should wait a longer amount of time than in normal conversation for appropriate turn-taking, as users may take more time to reflect on conversational well-being exercises.
} & {Acknowledgement and active listening were important for users to feel listened to (Section \ref{sec:convergence_capabilities}). Because verbal adaptation is limited (guideline 1), backchanneling and appropriate turn-taking can address these needs. Backchanneling has been shown to increase perception of active listening \cite{jones2011supportive, motalebi2019can}.
}  \\
\midrule
    {\textbf{(R4)} Verbal adaptation should be limited to preserve well-being practice efficacy
} & {We propose adaptive verbal utterances should be limited to portions of the interaction that do not affect the well-being practice (e.g. introductions, greetings).
} & {Participants call for extensive verbal adaptation in order to feel acknowledged (Section \ref{sec:convergence_capabilities}). However, well-being practitioners say extensive verbal adaptation (such as a conversation), could undermine efficacy of well-being practice (Section \ref{sec:divergence_emotional_adaptation}, \cite{steinbrecher2020providing}). This is a conflict in preferences of coaches and participants. The guideline aims to preserve the efficacy of the well-being intervention and satisfy user requests for verbal adaptation. }    \\

  \bottomrule
\end{tabular}
\end{table}

\begin{table}
  \caption{Design recommendations rationale}
  \label{table:design-guidelines-rationale2}
  \begin{tabular}{A{2,5cm} A{4,5cm} A{5,9cm}}
    \toprule
      {\textbf{Recommendations (HOW)}} & {\textbf{Issue (WHAT)}} & {\textbf{Rationale for recommendations (WHY)}}  \\
    \toprule
    \multicolumn{3}{l}{\textbf{Category: Ethical considerations}} \\
    \midrule
    {\textbf{(R5)} Safeguarding and risk assessment are required prior to HRI} & {A risk assessment needs to be completed before interaction with the robot, in order to verify that users are not in a mental health crisis. Such users should be directed to other, more appropriate resources.
    } & {Participants and coaches noted that a robot would not be appropriate for a person in a mental health crisis (Section \ref{sec:convergence_ethical}). The proposed robotic well-being coach is not to be used with people who need clinical mental health interventions \cite{fiske2019your, stahl2016ethics}.
    }    \\
    \midrule
     {\textbf{(R6)} Data collection needs informed consent to preserve privacy} & {Participants should be informed on what data is being collected and why, how it is stored, and how it is used to adapt or personalize the interaction.} & {Participants call for adaptation and personalization across all studies (Section \ref{sec:convergence_ethical}), but are also concerned about data collection and privacy (Section \ref{sec:divergence_adaptation_privacy}). Participants note that all users should be informed of data collection, following GDPR \cite{voigt2017eu}.} \\
    \midrule
    {\textbf{(R7)} Researchers should educate users keeping in mind emotional considerations and user characteristics} & {In addition to educating users about the collection of their data and privacy, users should be informed accurately of the capabilities and limitations of the robot. The education should be given \hl{with the content and flexilibity that participants with varying levels of technical literacy can understand it}.
    } & {User characteristics (e.g., level of experience with technology) may shape the language needed to make user education understand. While the robot being non-judgemental was an advantage across the studies (Section \ref{sec:convergence_advantages}), participants should still be aware of what information they give to the robot (Section \ref{sec:convergence_ethical}), which is especially important in applications with vulnerable users or data \cite{riek2014code}, \hl{and how this data is used. Participants should also be informed that a robot with empathic expression  does not actually ``possess'' empathy as a human does.}
    }    \\
  \bottomrule
\end{tabular}
\end{table}

\subsection{R1 - Robot form should follow its function}
\label{sec:guideline_form_function}

The robot's form, i.e. its appearance and embodiment, should be of a similar level of sophistication to its functionality. In Study 1 (Section \ref{sec:study1}), participants had conflicting opinions on whether they would prefer a humanoid or an abstract robot (see Section \ref{sec:divergence_form}). However, participants noted that they would expect the robot's appearance to match its level of communication (i.e. they wouldn't expect a robotic dog to speak, but they would expect conversation abilities from a humanoid robot such as Pepper). As such, we utilised the humanoid robot Pepper in the Positive Psychology exercises (Study 2, Section \ref{sec:study2}) and in the demonstration of Positive Psychology in Study 3 (Section \ref{sec:study3}). Pepper received mainly positive feedback in Study 2, whereas robot form was not discussed in Study 3. 
Previous works also provide evidence of the relation between the robot's form and user expectations. For example, \citet{haring2018ffab} claimed that when interacting with robots people are biased by visual perception, and they attribute a degree of functionalities to the robot, taking a cognitive shortcut. They showed how this bias can affect the design of a robot and discuss the implications for HRI. Also, \citet{goetz2003matching} showed that a match between the appearance of the robot and its functionality can improve the collaboration during a human-robot task. In \cite{hosseini2017love}, the authors conducted a survey with 56 participants who were asked to assess seven different robotics platforms. Their results showed that participants scored the humanoid robots as most useful. 

If designers want to convey conversational capabilities for a robotic well-being coach, we recommend that robots with a more humanoid appearance are chosen. In the case of Mindfulness and Positive Psychology, if these exercises are not administered in a conversational format, abstract or even animal-like robots may be applicable.

\subsection{R2 - Robot voice should emphasize variable prosody and slow pace}

The use of speech for robots to communicate with humans is a key factor for a successful interaction. While the voice of a robotic well-being coach was not extensively discussed during any of the studies, some recommendations can be made. Pepper's default voice received mainly positive feedback during Study 2 (Section \ref{sec:study2}). However, in Study 3 (Section \ref{sec:study3}), participants noted that a robot's voice may be too monotonous during longer instructional utterances, that introduce the well-being exercises to be conducted. A monotonous voice may result very difficult to pay attention to (as discussed specifically in relation to the Miracle Question, which is a method used in SFP). 

Within the HRI literature, past works have widely explored the effect of robot's voice in terms of prosody and pace, especially to convey emotions via speech \cite{crumpton2016survey, winkle2017investigating, jee2010sound}. For instance, \citet{crumpton2015validation} investigated the use of an open source speech synthesizer to convey emotions through the robot's prosody. Their results showed that participants were able to distinguish the emotion conveyed by the robot using its voice. Also, several studies have demonstrated how the robot's voice affect the user perception of the robot's personality \cite{aylett2019right, mou2020systematic, dou2019effects}. For example, \citet{dou2019effects} conducted a study with 15 students and a Pepper robot to analyze the correlation between the perceived personality of the robot the its voice. Their findings showed that the robot with the child-like voice was more likely to be perceived as extroverted, passionate, and relaxed.

We suggest that especially if the robot will be giving long instructions, special attention should be paid to making the robot's voice human-like with varied prosody, and that emphasis is placed on important words in the robot's utterances. This becomes especially important in well-being practices such as Mindfulness and Meditation, where the coach often talks for longer periods of time without input from the coachee, while it conducts guided meditation. During Mindfulness and Meditation sessions, the robotic coach should also have a slow and calm pace of speech, in order to encourage a meditative experience during the practice.

\subsection{R3 - Acknowledgement and active listening should be implemented through backchanneling and appropriate turn-taking}
\label{sec:backchanneling}

Due to the difficulty in creating a sense of the robot \emph{actively listening} (which the participants called for, see Section \ref{sec:convergence_capabilities}) through \emph{verbal adaptation}, we suggest that robotic well-being coaches do this through \emph{backchanneling} via \emph{body language} and \emph{phatic expressions}. Phatic expressions (such as ``Uh-huh'' and ``Mm hmm'') can be used to convey the active listening of the robot while the user is speaking. Backchanneling has been shown to improve the functioning of human-robot teams \cite{jung2013engaging}. Well-timed movements such as nodding the robot's head when the user is talking have also been previously shown to improve the user's impression of the robot listening to them \cite{park2017telling, murray2022learning}. Recently, data-driven methods have been used to create believable emotional robotic gestures based on an initial set of hand-designed gestures, in order to introduce a variety of expressions that contributes towards a robot's life-likeness \cite{marmpena2022data}. Such well-timed and believable body language was requested by participants across all three studies.

We suggest that roboticists create robotic well-being coaches carefully designed with appropriately timed backchanneling cues that the robot expresses, especially during conversational well-being practices such as SFP or Positive Psychology. 

Our study results show that appropriate timing of robot expressions emerges as being relevant to the smoothness of turn-taking as well. The timing of the robot taking its turn to speak (as expressed via verbal utterances and gestures) should be carefully considered as well. Turn-taking has been previously predicted in human-robot interaction with the observation of backchannels and fillers \cite{hara2018prediction}. Additionally, multimodal perception has been been used to improve the detection of a human's continuing speaking turn by observing their filled pauses and gaze. This significantly reduced robot interruptions, and increased the time people spent speaking to the robot \cite{bilac2017gaze}. In Study 2 (Section \ref{sec:study2}), participants commented on how they weren't sure when they should be talking, and when it was the robot's turn. This led to the participants sometimes experiencing the robot interrupting their stream of thought, reducing the experienced efficacy of their self reflection. 

We recommend that designers of robotic well-being coaches leave sufficient time between the participant finishing their speaking turn and the robot beginning its next utterance. These times may be longer than would be considered natural during an every-day conversation, due to well-being exercises sometimes requiring the participant to spend more time reflecting on their experience, and thus resulting in longer pauses in the participants' speech. The robot should also be able to discern between the participant being silent and ending their speaking turn, and recognize users using phatic expressions (such as ``Hmm'') to indicate that they are still continuing their speaking turn after the pause.

\subsection{R4 - Verbal adaptation should be limited to preserve well-being practice efficacy}

Participants often called for extensive adaptation of the robot's utterances and word variety, and for the robot to personalize its utterances based on what the participant had said (Section \ref{sec:convergence_capabilities}). However, as seen in all three studies, coaches may disagree with this on the basis that it might jeopardize the efficacy of the well-being practice (see Section \ref{sec:divergence_emotional_adaptation} for further discussion). In the case of SFP, the structure and the specific phrasing of sessions is carefully structured, and interfering with it could disrupt the clinically-proven efficacy of the sessions. For example, the Miracle Question, a tool used in SFP, uses specific phrasing such as ``imagining'' a miracle occurring, which encourages coachees to think outside the box \cite{steinbrecher2020providing}. The same concern for preserving efficacy also applies to participants calling for the robot to remind them of what they had said in previous sessions (i.e. creating a memory for the robot) which  is not normally done in SFP, as the practice underlines that the coachee should determine important topics to discuss themselves. The coach was of the opinion that the robot could remember what content had been dealt with in previous sessions (i.e. session topics), but not volunteer to the participant what they had previously said, unless they brought it up themselves. In Study 2 (Section \ref{sec:study2}), the coach reviewing the script determined that minimal verbal acknowledgement to the participant's utterances is acceptable, but that the practice has clinical reasoning behind it, and that extensive alterations to approved phrasing may risk efficacy. 

In the case of well-being practices based on Mindfulness, the Meditation/Mindfulness coach (study 1, Section \ref{sec:study1}) underlined that the ability to ``read the room'', and give instructions in the moment as needed, is important. Meditation has previously been flexibly adapted to fit different needs. For example, walking meditation was adapted to older adults by practicing walking meditation at a slower pace, and always with an aid available to support possible loss of balance \cite{morone2014adapting}. This flexibility was also emphasized by the Mindfulness coach in Study 3 (Section \ref{sec:study3}), who noted that the robot should emphasize each participant's specific \emph{strengths}. This ability of the robot to ``read the room'' and give instructions as needed might need the robot to have sophisticated sensors in order to discern people's level of calmness during individual or group meditation (e.g. from fidgeting, breath rate, etc), and determine when it is appropriate to intervene. Alternatively, this problem could be addressed with a robot intervening at set intervals, even if this does not reach the level of human Mindfulness coach performance. In the case of Mindfulness and Meditation, too frequent interventions could begin to disturb the participant's concentration, thus lessening the efficacy of the practice.

Adaptation is a difficult problem, as users expect a robot, especially a humanoid robot, to adapt to them and \emph{acknowledge} them. However, adapting well-being practices, which often are based on thoroughly tested clinical interventions, can be a difficult and even risky endeavour. The level of adaptation should be based on how appropriate it is for each intervention. Here, we conclude that based on the aforementioned data, SFP should not be extensively adapted, and Mindfulness should be adapted somewhat. As the SFP coach said in Study 3, they would prefer it to have minimal adaptation, to minimize risk of responding incorrectly. This coach noted that the level of improvement in the users' interaction experience by using verbal adaptation may not be worth the risk in the case of failure. Thus, we recommend that roboticists create a ``sense'' of adaptation by introducing \emph{wording variety} in certain statements that do not affect the procedure of the well-being practices (such as introductions, explaining the content of the sessions, etc.), but not during the well-being content itself.

\subsection{R5 - Safeguarding and risk assessment are required prior to interactions} 
\label{sec:safeguarding}

Another concern of participants and coaches across studies was the safeguarding and risk assessment of participants (Section \ref{sec:convergence_ethical}). Recently, HRI literature has been giving more attention to ethical considerations and how it is ethically and morally required to conduct a prior assessment of mental health of users interacting with the robot to safeguard their health \cite{fiske2019your}. \citet{fiske2019your} concluded their paper acknowledging that further research is needed to address the broader ethical and societal concerns of robotic technologies. Also, \citet{stahl2016ethics} discussed ethical aspects, and they proposed a new framework to evaluate ethical principles and support researchers in this process, namely Responsible Research and Innovation for healthcare robotics. 

Analogously, the findings of Study 3 (Section \ref{sec:study3}) suggest that a robot is not equipped to address such situations, and we object to any attempts to do so, as this may pose a danger to the user. While we argue that a robotic mental well-being coach could be useful for non-clinical populations wishing to maintain or improve their well-being, we do not want to see such robots replacing or even stepping in for human well-being professionals in situations where well-being is a matter of clinical intervention. We emphasize the distinction between coaching (focusing on the present and the future, and the flourishing of the coachee) and therapy (more focus on the past, and addressing mental health disorders) \cite{hart2001coaching}, where a robotic mental well-being coach can be appropriate for the prior but should not be performing the latter.

\hl{Currently, robots do not have adequate capabilities to identify a situation where the user's well-being might be at risk, and as such pre-screening needs to be carried out by a  human. If risk to a participants safety arises despite pre-screening, it is important to have a safeguarding procedure established and communicated to the participant beforehand, which will be carried out depending on the context. For example, in the context of a research study, researchers who are not psychologists and not in a clinical relationship with a person will not be able to provide mental health care. Instead, directing participants to appropriate resources with pamphlets and other informational sources can be appropriate (see literature on the impact of pamphlet distribution, e.g.,} \cite{vetto1996impact, ashizawa2018use}). \hl{These limitations of the robot and the researchers should be communicated to the user beforehand: in case they experience discomfort or safety concerns during their interaction with the robot, they can take a pamphlet with them. 

In contexts where a robotic coach might be deployed outside of a research setting and it did have the capability to detect safeguarding situations, e.g. a university, a workplace, or at home, appropriate safeguarding needs to be designed on a case-by-case basis. Additionally, the user should be informed and agree to safeguarding procedures beforehand. For example, at home, the user may prefer to use the robot without the robot engaging in safeguarding, as one would use a diary. In other contexts, e.g. at a university, it might be important to alert university healthcare services in potential safeguarding situations. However, such case-by-case recommendations are out of the scope of this paper, and should be researched further in each context, and as robotic safeguarding situation identification capabilities evolve.}

We suggest that \hl{in the context of current robot capabilities and within research settings}, participants be pre-screened by a human before beginning the use of a robotic mental well-being coach, in order to identify situations where a prospective user may be in significant distress or suffering from a mental health crisis.

\subsection{R6 - Data collection needs informed consent to preserve privacy} 
\label{sec:ethicalPrivacy}

The ethical consideration that received most attention throughout all three studies was data collection and privacy (Section \ref{sec:convergence_ethical}). Previous research has also placed emphasis on the preservation of privacy in digital well-being applications. Privacy control approaches have been detailed with a user-centric approach, especially for well-being applications that collect personalized data with ubiquitous sensors in environments such as the user's home or workplace \cite{bokhove2012user}. Some virtual coaching applications have opted for,  e.g., not collecting demographic information to preserve privacy \cite{jeannotte2021time}.

Our findings show that the consideration for privacy was in conflict with the preference for the robot's \emph{personalization and adaptation}, especially during the group discussion in Study 3 (Section \ref{sec:study3}). Specifically, the Mindfulness coach called for adaptation by the robot to a specific user's strengths, but did not want the robot to collect any personal data. Though efforts were made by the researchers to explain why both these things could not be achieved, this concept was not necessarily understood by all participants due to lack of technical knowledge. This indicates that while PD strives to elicit design choices from experts and prospective users of robots, these choices may not always be realistic.

In contrast, the SFP coach present in Study 1 (Section \ref{sec:study1}) noted that privacy may be increased when interacting with a robotic coach in comparison to a human coach. The coach noted that during clinical interventions, information about patients is shared with other clinicians (patients are made aware of this). While we do not propose the use of a robotic well-being coach for clinical populations, this presents and interesting case where a robot may in fact preserve privacy, even though it is collecting information to be stored in a more accessible format than a human coach. 

We recommend that the data collection and processing performed by a robotic well-being coach should be efficiently communicated to its users, following GDPR. This is especially important as users may have differing levels of technical knowledge, and may not perceive that any information is being recorded unless specifically stated. \hl{Users should be informed about the type of data that is being collected, e.g., the content of their utterances, their facial expressions, or their gestures. Users should also be informed and give consent to how this data will be stored and for how long --- as is stipulated by GDPR. As discussed in Section} \ref{sec:safeguarding}, \hl{it should also be explained whether and how participants' data will be used for safeguarding purposes in their particular context.} To preserve \emph{transparency} of the robot's operation, users should also be made aware how the robot's behaviour (e.g. selection of utterances) may be affected by the collection of their data.

\subsection{R7 - Researchers should educate users keeping in mind emotional considerations and user characteristics} 

The topic of user education encompasses different concerns that we have identified in our studies, even if not explicitly mentioned by the participants. This ethical consideration aims to address varying \emph{user expectations}, and accurately convey to the user the \emph{capabilities and limitations} of a robotic well-being coach. This involves educating users by giving them an appropriate impression of the robotic coach's capabilities and limitations, especially in the case of users that are less technologically knowledgeable. This is not a simple matter, as emphasizing the robot's limitations too enthusiastically may reduce the effectiveness of well-being exercises being conducted.  Especially in applications with vulnerable users, such as people disclosing details about their well-being, should be given accurate information about the operation of the robot due to its emotional proximity \cite{darling2015s, darling2016extending}.

\emph{Emotional consideration}, i.e., the consideration of how the emotional relating of users to the robot may influence their behaviour \cite{axelsson2021participatory}, is important when designing of robotic well-being coaches. In Study 1 (Section \ref{sec:study1}), the SFP coach noted that when they conduct clinical therapy, they may notice patients developing an \emph{over-reliance} on the therapist, idealizing them or viewing them as the only person that would understand them. While this may not be a big concern in non-clinical well-being coaching, it should still be considered when designing such robotic well-being coaches. Such over-reliance can be mitigated by informing the users that the robot is intended to be used as a tool. 

\hl{A concern in robot design has been that the design of robot capabilities, such as empathic expressions, can be seen as deceptive or inauthentic} \cite{westlund2015deception, boulicault2021authenticity} \hl{However, previous research has also found that expressions of empathy can improve people's experiences with robots} \cite{niculescu2013making}, \hl{and robotic coaches in particular} \cite{axelsson2022participant, jeong2023robotic}. \hl{We recommend that researchers be aware that while participants express their desire for a robot to have empathic expressions (see Section }\ref{sec:empathy_expression}), \hl{and that those expressions can improve the well-being practice, it does introduce some concerns. In order to mitigate these concerns, if the robot uses empathic expressions, we recommend that at the beginning of a study (or if this is not feasible, at the end of a study), researchers have an in-person discussion with participants. This discussion should inform users that the robot has been designed to express empathy in order to facilitate a better coaching experience} \cite{niculescu2013making, axelsson2022participant, jeong2023robotic}\hl{, but this does not however mean that the robot ``possesses'' empathy in the human sense. Informing users of robotic capabilities can be helpful in clarifying misconceptions and mitigating feelings of deception} \cite{westlund2015deception, boulicault2021authenticity}.

User education is also important in the case of users lacking \emph{technological knowledge} (e.g., elderly users, as suggested by a participant to be a user group to specifically benefit from a robotic coach in Study 2, Section \ref{sec:study2}) to accurately judge the amount of \emph{privacy} that a robotic coach may afford them (as discussed in Section \ref{sec:ethicalPrivacy}). Previous studies have noted that especially in the case of vulnerable users or sensitive applications, special attention should be paid to privacy \cite{axelsson2021participatory}. This concern can be addressed by informing users about their data being collected and processed in comprehensible language, either before beginning any sessions with the robot coach, or prior to each session. \hl{We recommend that at the beginning of the sessions, a person knowledgeable about the robot (e.g., the researcher) be available to the participants to answer any questions. The researcher should clarify what the robot is currently capable of and is ``good at'' (i.e., facilitating non-clinical well-being exercises), and what it is not capable of and is ``bad at'' (i.e., identifying safeguarding situations or conducting meaningful therapeutic conversations). This way, the participant can form an accurate conception of the robots capabilities and limitations, and can clarify any technical terms or request further explanation from the researcher, if their technological knowledge is a barrier to understanding how the robot works.} This especially becomes a concern in the case of commercially available robots, where the informed consent of users may not be as meticulously implemented as in academic studies.

We recommend that designers of robotic well-being coaches take user education into consideration on a user-by-user basis, and provide user education as applicable to their specific application keeping in mind emotional considerations and user characteristics.

\section{Conclusion, Limitations and Future Work}
\label{sec:concl}
In this paper, we present the \textcolor{black}{qualitative analysis} of three studies focusing on the design and evaluation of robotic mental well-being coaches. Based on this analysis, we present design guidelines including ethical considerations for such robots. Additionally, we discuss what aspects of the design of such a robot should be considered in the future. We present these analyses and recommendations in order to help future designers and roboticists creating robotic well-being coaches, and to operationalize the knowledge we have found from these studies.

We propose the design recommendations and guidelines, as well as the ethical considerations presented in this paper based on the coverage of the three studies, which examined a robotic well-being coach from different perspectives (see Table \ref{table:study-comparison}) and achieved data saturation (see Section \ref{sec:methodology}). However, our studies did not examine the longitudinal use of robotic mental well-being coaches (though the motivations to use such a robot in the long-term were discussed in Study 3, Section \ref{sec:study3}). 
Future research should examine how robots designed with these considerations in mind are experienced by participants in the long term, over multiple interaction sessions. Such examinations of robots could potentially bring forth the need for additional recommendations regarding capabilities that can only be examined in the long term, such as long-term personalization. Additionally, \hl{our paper describes an iterative design process to distil recommendations to inform the design and development of robotic coaches for mental well-being limited to an adult target population. The design and ethical recommendations proposed do not address the needs of more vulnerable populations such as children, the elderly, or people with neuro-developmental disorders. Future research should be conducted for designing robotic coaches for such target populations.}
There are several aspects that are known to impact HRI we have not discussed in this work and could be addressed in future works. We highlight them as follows.

\textbf{Cultural Influences.}
In recent years, researchers have argued the importance of considering different cultures during human-agent interactions \cite{joosse2014cultural}; even so, most of the HRI works has not yet incorporated those aspects, designing robots that are still mainly influenced by Western cultures. 
The few works that focused on HRI cross-cultural research investigated the Western and Eastern cultures represented respectively by the American and Japanese populations \cite{lim2021social}. However, even within the same culture (e.g., Western), many differences and nuances are observed (e.g., Italians vs. British people) \cite{conti2015cross}.
In our context, cross-cultural research would be beneficial for creating robotic well-being coaches. Future work could examine how the guidelines presented here are applicable to different cultures and populations, expanding the results of Study 3, with participants remarking that e.g., a user's socio-economical status might affect how they experience the robot. 

\textbf{Role of Context.}
Lately, the HRI community has highlighted the role of the context during human-robot interactions \cite{cameron2015framing}. In fact, human behaviors, mental processes, and emotional states vary depending on the environment in which they occur \cite{salam2015multi}.
In parallel, many efforts have been made to perceive and model the social and normative context in human-agent interactions within the affective computing field \cite{wang2018deep, de2022vision}. It is crucial to develop robots that are context-aware to be able to adapt appropriately to various situations during human-robot interactions. Particularly, in our work, designing robotic coaches that are aware of the context could positively impact the well-being practice. Future work could investigate further what is the role of the context in this application scenario. 

\textbf{Cross-fertilization Between Affective Computing and Social Robotics.}
The affective computing field is advancing in the automated analysis of emotional and social signals. However, social robotics has not yet integrated these most recent developments into robotic platforms \cite{celiktutan2018computational}.  In addition, due to the necessity for real-time processing skills and the lack of computational capacity of the robotic platforms on the market, deploying social robots in real human-robot interaction scenarios is still an open challenge \cite{park2020towards}. Future work should focus on the collaboration between the affective computing and social robotics fields by leveraging cloud computing or external sensors as suggested in \cite{park2020towards}, to design autonomous robots that can deliver well-being exercises.

\begin{acks}

We would like to thank Dr. Indu Bodala for contributing to Study 1, Nikhil Churamani for contributing to Study 2, and Katherine Parkin for her contributions to Study 1, Study 2 and Study 3. We thank Dr. Milo Phillips-Brown for the insightful conversation on the ethics of robotic coaches. We would like to thank all study participants and coaches for their valuable contributions to this research.

\textbf{Funding:} M. Axelsson is funded by the Osk. Huttunen foundation and the EPSRC under grant EP/T517847/1. M. Spitale and H. Gunes have been supported by the EPSRC/UKRI under grant ref. EP/R030782/1 (ARoEQ).
\textbf{Open Access:} For open access purposes, the authors have applied a Creative Commons Attribution (CC BY) licence to any Author Accepted Manuscript version arising.
\textbf{Data Access Statement:} Overall qualitative analysis of research data underpinning this publication is contained in the manuscript, as well as manuscripts related to the original studies as referred to in the manuscript. Additional raw data related to this publication cannot be openly released as the raw data contains and transcripts of the participants’ interviews, group discussions, and interaction with the robot, which were impossible to anonymise.
\end{acks}

\bibliographystyle{ACM-Reference-Format}
\bibliography{main}

\appendix


\includepdf[page={1}, angle=90,  pagecommand={\section{List of codes}, \label{app:codebook}} ]{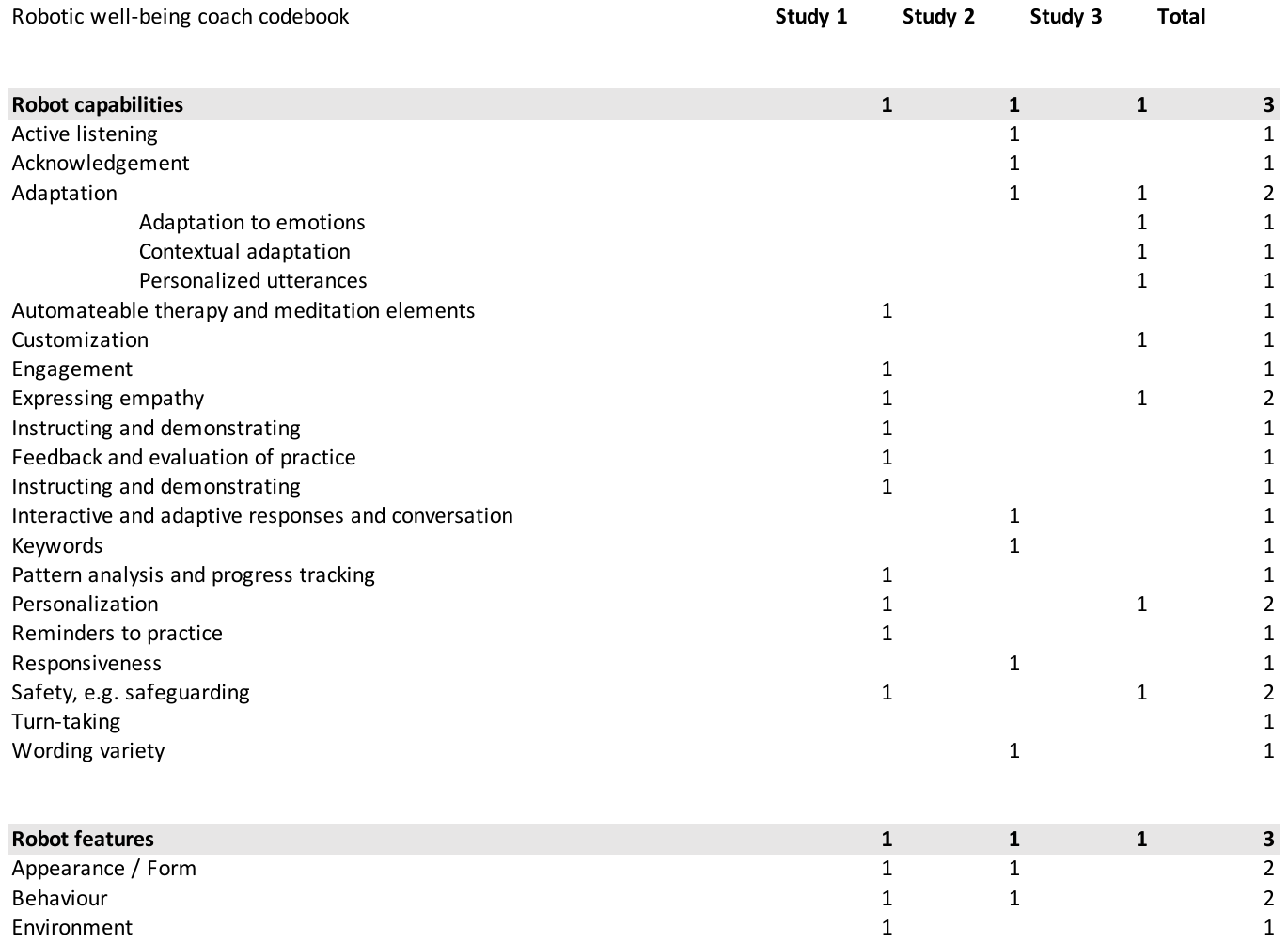}
\includepdf[page={2}, angle=90 ]{THRI_codebook-2}
\includepdf[page={3}, angle=90 ]{THRI_codebook-2}
\includepdf[page={4}, angle=90 ]{THRI_codebook-2}
\includepdf[page={5}, angle=90 ]{THRI_codebook-2}


\section{Study 1: Prospective user interview}
\label{app:study1-user-interview}

An example prospective user interview. We modified interviews according to the participant's level of active well-being practice: currently doing well-being practices, considering doing well-being practices, and previously did well-being practices but stopped. Additionally, we tailored interview questions based on participants' responses to a survey prior to the interview. This example for the semi-structured interview is for a participant who is currently doing well-being practices.

\begin{enumerate}
    
    \item Current well-being practices
    \begin{itemize}
        \item You said you are currently doing well-being practices. Could you say what makes you motivated to do them?
        \item Are you experiencing particular benefits? In your survey you specify reduced anxiety, improved performance, better sleep, developed gratitude, increased happiness, and reduced stress. Could you elaborate on this?
        \item Are there particular challenges you are addressing with well-being practices?
    \end{itemize}

    \item Questions about specific survey answers
    \begin{itemize}
        \item In the survey you say you are currently doing mindfulness and meditation, what made you decide to do these things? Had you seen or heard of them before? In your survey you specify personal interest and research.
        \item You said you are aiming to achieve reduced anxiety and improved performance and better sleep and develop gratitude and increase happiness and reduce stress with the practices. Could you elaborate?
        \item You said that you are practicing once a week. What motivates you to keep practicing at this frequency?
        \item You've practiced since \emph{[year omitted]}. What made you stick with it? 
        \item In your survey you specify having a routine to your practice. Could you elaborate on that?
    \end{itemize}

    \item Tools used for well-being practice
    \begin{itemize}
        \item You say you are using non-digital tools: note taking by hand and in-person practice with an instructor. Could you elaborate on how you’re using them, and when?
        \item What benefits are you finding with these tools?
        \item Are these benefits motivating you to practice, and to stick with the practice?
        \item You say you were previously using digital tools. Could you tell me a bit about them?
    \end{itemize}

    \item Technology providing help
    \begin{itemize}
        \item You mention that you stopped using technology during your practice because it was inconvenient and you were not satisfied with the quality of the technology and it felt impersonal. How did you experience these things?
        \item Did you experience any benefits when using technology?
        \item Did you experience or could you imagine experiencing any benefits to using technology for your practice?
        \item Are there any additional features that would make you want to use technology for your practice again?
        \item Is there something that would make you stop using technology again?
        \item Did you have any concerns about using technology for well-being practice?
        \item Have you seen or heard about some interesting well-being practices employing technology in some way?
        \item Do you have ideas about how technology could be used to create well-being practices?
        \item Does something else come to your mind about technology and well-being practices?
    \end{itemize}

\end{enumerate}

\section{Study 1: Coach interview}
\label{app:study1-coach-interview}

An example semi-structured coach interview structure, here for the Life Coach. All interviews shared a general structure, and were then tailored specifically to the coach. Some questions were based on background information the coach had given prior to the interview. Prior to questions about robots, all coaches were shown the same five photos and videos of robots used with the prospective users in the group discussions.

\begin{enumerate}
    
    \item Types of practices
    \begin{itemize}
        \item What types of coaching practices do you instruct?
        \item Do you do one-to-one or group instructions? If both, what are the differences?
        \item Could you briefly expand on the concept of Life Coaching?
        \item You mentioned you use positive psychology as part of coaching, would you please expand on that? How do you use PS in the context of life coaching?
        \item You mentioned that you use cognitive behaviour psychology as part of coaching, would you please expand on that? How do you use cognitive behaviour psychology in the context of life coaching?
        
    \end{itemize}
    
    \item Benefits and goals
    \begin{itemize}
        \item What do you think are the benefits of life coaching for the coachee ?
        \item What have your coachees told you they think are the benefits?
        \item What goals do the coachees  usually have?
        \item How are these goals identified?
    \end{itemize}
    
    \item Coachees' motivations
    \begin{itemize}
        \item What types of needs do people have when they come for coaching?
        \item How do the coachees find out about your services and life coaching? Do they come through recommendations / word of mouth?
    \end{itemize}
    
    \item Practicalities
    \begin{itemize}
        \item How long are coaching sessions usually?
        \item How often are coaching sessions usually?
        \item How many sessions are there usually?
    \end{itemize}
    
    \item Tools and technology
    \begin{itemize}
        \item Do you use tools when you instruct / coach? What kinds of tools?
        \item How are the tools helping the coachee?
        \item How are the tools helping you as the coach?
        \item Do you use technology as a tool when you coach? What kinds of technology?
        \item How is technology helping the coachee?
        \item How is technology helping you as the coach?
        \item What advantages and disadvantages do you see practicing online virtually, vs. face-to-face?
        \item What are the differences when the sessions are mediated through technology vs. having face-to-face sessions?
        \item Do you think there could be benefits to using technology during COVID-19?
    \end{itemize}
    
    \item Further technology
    \begin{itemize}
        \item How do you think technology could help further with coaching?
        \item Are you aware of technologies that have helped people with their coaching or their goals as coachees? (e.g. Headspace)
        \item Have you ever used such technologies?
        \item What do you think the benefits of such technologies are?
        \item Do you think such technologies can become the primary method of delivery for coaching? Or rather, are they secondary and should be used in support with a human instructor?
    \end{itemize}
    
    \item Specific questions about practice
    \begin{itemize}
        \item How do you decide which goals to work on with clients?
        \item Who is in charge of the coaching sessions you instruct? Who leads the interaction?
        \item How do you decide when to talk and when to stay silent?
        \item Do you use a script during the sessions? 
        \item How do you adapt the sessions? According to the day? According to the person?
    \end{itemize}
    
    \item Robots
    \begin{itemize}
        \item Would you be able to imagine such a robot providing or giving a life coaching session?
        \item Which aspects of life coaching do you think would be appropriate in this context, and why?
        \item Are there any other coaching practices that you think would be appropriate to be delivered by such robots? Why?
        
    \end{itemize}
    
    \item Robot advantages and disadvantages
    \begin{itemize}
        \item What do you think the advantages of using a robot for these types of practices may be? Previous participants have mentioned for example: personalization and adaptation, analyzing previous practice patterns and progress, accessibility (having it in your house/work), consistency, and being non-judgemental.
        \item How about the disadvantages? Previous participants have mentioned for example: Privacy, expensiveness, repetitive interactions, unreliable technology, and intrusive reminders.

    \end{itemize}
    
    \item Robot capabilities
    \begin{itemize}
        \item Some participants noted they like self-disclosure done by their coach --- do you use self-disclosure (i.e. sharing your own experiences) as part of coaching instruction?
        \item Do you think you express emotions during sessions? How?
        \item Do people need guidance during the coaching sessions?
        \item Do people get sidetracked when exploring an area or a particular topic?
        \item Do people need to be given some time to think?
    \end{itemize}
    
\end{enumerate}

\section{Study 2: Post-interaction interview}
\label{app:study2-interview}

The questions asked during the semi-structured post-interaction interview:
\begin{enumerate}

  \item Robot
  \begin{itemize}
    \item What do you think was good about the robot’s appearance? What did you like?
    \item What do you think was bad about the robot’s appearance? What would you change?
    \item How appropriate do you think the robot was for helping you focus on the positive aspects in your life? How appropriate do you think the robot was for helping you foster a positive attitude toward things that happen in your life?
    \item How appropriate do you think the robot was for these coaching tasks?  
    \item How useful do you think the robot was for these coaching tasks? 
    \item How beneficial do you think the robot was for these coaching tasks?
  \end{itemize}
  
  \item Behaviour
  \begin{itemize}
    \item What do you think was good about the robot’s behaviour? What did you like?
    \item What do you think was bad about the robot’s behaviour? What would you change?
    \item How appropriate do you think the robot’s behaviour was for these coaching tasks?  
    \item How useful do you think the robot’s behaviour was for these coaching tasks?  
    \item How beneficial do you think the robot’s behaviour was for these coaching tasks?  
    \item Do you think the robot understood what you said?
    \item Do you think the robot understood how you felt?
    \item Do you think the root adapted to what you said and did?
    \item Did you notice any differences?
      
  \end{itemize}
  
\end{enumerate}




\section*{Supplementary material}

# E    SFP SCRIPT

## 1. Introduction

Robot: Hello, My name is X. What should I call you?

Human: NAME

Robot: Nice to meet you NAME

Robot: So we've got 4 sessions together, each 20 minutes long, with about a week between each session. We're going to be using a Brief Solution-Focused approach. It's a type of conversation that helps people think about what their best hope and goals for the future are, and think about what strengths and resources they have as a person. And how they might know that they are getting closer to those best hopes, and noticing signs of progress towards these best hopes, just through conversation. Noticing what works and doing more of it.

Robot: And what we find is that when we talk about people's best hopes and signs of progress, that they naturally start noticing what works well, and doing more of what works well, without really trying. That's kind of the approach. Do you have any questions about that? Please say 'yes' or 'no'.

Human: No.

{Human: Yes.

Robot: I will call in the experimenter to answer your questions. Experimenter, could you please come in?}

## Session begins

### 2.  What are your best hopes / what are you hoping to achieve

Robot: So thinking about what your best hopes are, how would you know that these sessions have been useful to you? What would you like to work on?

Human: I would like to have a better mood, I often feel sleepy.

Robot: So, you would like to have a better mood and feel less sleepy.

Human: Yes

## 3. Miracle question

Robot: Say you leave here today, and you do what you have to do for the rest of the evening, and you go to sleep. And something odd happens in the night, so during the night a miracle happens. And your best hopes are realized, that's what the miracle is. And you wake up the next morning, but you were asleep when the miracle happened, so you don't know that it happened. What would be the first signs that would tell that a miracle had happened? You don't know it did. What would you notice?

Human: I would feel excited about my day, and be able to exercise in the morning. I wouldn't feel miserable. I would get things done for the day, and follow my routine.

Robot: So you'd wake up, you'd feel excited, you wouldn't lie in, you would get going. You would follow your routine, what is your routine?

Human: I would go for a run in the morning, and go to work early. And I would do something I enjoy after work. And actually set time aside for doing it, and not feel tired. Have a nice dinner and go to sleep early.

Robot: Are there days where it felt like what you're describing?

## 4. Are there instances that are good, how do you notice?

Robot: Are there instances when you notice you have achieved your best hopes, in the past or the recent past? When you've woken up and done all the things you've said?

Human: Sometimes yes.

Robot: What percentage of days would you say that you achieve your best hopes?

Human: Maybe 5 days a week.

Robot: What happens on the days or instances where you achieve your best hopes? What leads to you achieving your best hopes on those days?

Human: I get up early, and I eat well.

Robot: So on the days when you achieve your best hopes, you get up early and you eat well?

Human: Yes and I notice I feel better on those days.

Robot: That makes sense.

Robot: Anything else that makes the days when you achieve your best hopes go better?

Human: If I see friends, I notice I feel better that day.

Robot: So you may have already noticed some signs of moving towards your best hopes. That often happens before people come to a session. They've already started working on the problem they're describing.

## 5. Scaling question

Robot: So where would you put yourself on the scale of 1-10 on the bad days? So 10 would be the miracle has happened/your best hopes have been realised and 1 would be 'you couldn't be further from this'.

Human: 2

Robot: And what about good days?

Human: 5

Robot: So you manage to sometimes get yourself from a 2 to a 5.

Human: Yes, on the better days.

Robot: So there are already things you are doing that can put you up on the scale.

Human: Yes.

## 6. First signs that things are shifting

Robot: What would be the first signs that things were starting to shift toward your best hopes?

Human: I would notice more of these days in the month.

Robot: Any other small things that you might notice?

Human: I think I would notice that I read my book more, since that's what I like to do when I'm not too tired.

## 7. Summary

Robot: So in summary, you said you would/wouldn't notice…
You would have more time…
Some signs this would already be happening are…
There has already been a change in the last few weeks where...

## 8. End of session

Robot: So we are coming up to the 20 minute mark. So my suggestion, every week will be very similar, so I will always suggest that you spend the week noticing signs, any small signs, that you're getting closer to your best hopes. So you don't need to do anything differently, just notice times when you feel you are achieving or feeling your best hopes/.

Robot: And it can be 30 seconds when you feel that way, or a whole week when you feel like that was a great week, like you feel at a 7 on the scale. It doesn't really matter the time, but it's just tuning in to those small moments. And I emphasize small signs, because people expect there to be something big, when actually it's the small things that kind of accumulate. A small sign can even be planning to do something according to your best hopes, even if you don't do it.

Robot: So next week I will ask you what has been better in relation to your best hopes. And if you keep an eye out over the next week, it makes it easier. Sometimes the best hopes can also be maintaining things as is for now. Does that make sense?

Human: yes.

Robot: The nice thing about solution-focused approach is that you're probably already doing things, you're just not noticing them. We're all doing things that work, but we're not paying attention. We tend to notice when things go wrong, but sometimes it can take a bit longer to notice when things go right. In solution-focused practice, we have a structured way to think through things, and having a conversation about it often helps. And the ideas might start to naturally come.

Robot: Thank you for the session.

Human: Thank you!

# 9. Long-term interaction – Example of Week 2

## Scaling question

Robot: So last time we talked about the miracle question, and how you would know that a miracle had happened over night. So what do you think, on the scale of 0 to 10, so 10 is that the miracle happened, and 0 is couldn't be further from that. So where would you put yourself on this scale this week?

Human: 6

Robot: And what makes you a 6 rather than a 5. How do you know you're a 6?

Human: I'm feeling less tired this week, I notice I can wake up earlier and I feel energized enough to exercise in the morning.

Robot: And is there anyone who would notice it first, if we think of someone who knows you really well, do you have an example in mind?

Human: My housemates have seen me around the house earlier in the morning.

Robot: Who do you think of the people who know you, would be the first to notice that you were maybe a 7 on the scale?

Human: I think my housemate, because they also wake up very early in the morning and we meet each other in the kitchen when I'm feeling more energized.

Robot: And what would you notice that told you you were a 7 on the scale?

Human: I think I would be even more energized in the mornings, and maybe prepare a better breakfast than the sandwich I'm having now. Maybe some eggs.

Robot: When you are able to do the things you want, how does that affect your mood?

Human: I feel I'm doing better and I feel more energized because of the better breakfast.



\end{document}